\documentclass[twocolumn, times]{aastex631}
\usepackage{CJK}
\usepackage{amsmath}
\usepackage{mathtools}
\usepackage{subfigure}
\usepackage[T1]{fontenc}

\defcitealias{PaperI}{Paper~I}
\defcitealias{PaperII}{Paper~II}

\def\equationautorefname#1#2\null{(#2)}

\begin{document}
\begin{CJK*}{UTF8}{gbsn}

\title{Radiation GRMHD Models of Accretion onto Stellar-Mass Black Holes: III. Near-Eddington Accretion}

\correspondingauthor{Lizhong Zhang (张力中)}
\email{lizhong4physics@gmail.com}

\author[0000-0003-0232-0879]{Lizhong Zhang (张力中)}
\affiliation{Center for Computational Astrophysics, Flatiron Institute, New York, NY, USA}
\affiliation{School of Natural Sciences, Institute for Advanced Study, Princeton, NJ, USA}

\author[0000-0001-5603-1832]{James M. Stone}
\affiliation{School of Natural Sciences, Institute for Advanced Study, Princeton, NJ, USA}

\author[0000-0001-7488-4468]{Shane W. Davis}
\affiliation{Department of Astronomy, University of Virginia, Charlottesville, VA, USA}
\affiliation{Virginia Institute for Theoretical Astronomy, University of Virginia, Charlottesville, VA, USA}

\author[0000-0002-2624-3399]{Yan-Fei Jiang (姜燕飞)}
\affiliation{Center for Computational Astrophysics, Flatiron Institute, New York, NY, USA}

\author[0000-0003-2131-4634]{Patrick D. Mullen}
\affiliation{Michigan SPARC, Los Alamos National Laboratory, Ann Arbor, MI}
\affiliation{Computational Physics and Methods, Los Alamos National Laboratory, Los Alamos, NM}

\author[0000-0001-7448-4253]{Christopher J. White}
\affiliation{Center for Computational Astrophysics, Flatiron Institute, New York, NY, USA}
\affiliation{Department of Astrophysical Sciences, Princeton University, Princeton, NJ, USA}

\begin{abstract}
We present a comprehensive analysis of four near-Eddington black hole accretion models from GRMHD simulations with full radiation transport.  This study investigates the dynamical effects of magnetic field topology and black hole spin using two representative choices of each.  Two stable near-Eddington solutions emerge in these models: a thin thermal disk embedded within a magnetic envelope when sufficient net vertical magnetic flux is present (e.g., vertical field $\gtrsim 5\times10^5$~G at $20r_g$), and a magnetically elevated disk when the net vertical flux is weak or absent.  One model initialized without net vertical flux evolves into the thin disk solution, as strong, anisotropic radiation feedback at high accretion rates promotes the accumulation of vertical magnetic flux in the inner disk.  In the thin thermal disk, accretion is driven primarily by mean-field Maxwell stress and proceeds largely within the magnetic envelope, while heat dissipation is spatially decoupled and concentrated near the midplane.  However, in the magnetically elevated disk, accretion occurs throughout the disk body and is comparably driven by mean-field and turbulent stresses; heat dissipation therefore occurs locally through turbulence.  Radiation transport is diffusion-dominated, enabling efficient radiative cooling ($\sim$4-10\%).  An optically thin wind is launched from the disk surface by combined radiative and magnetic forces, with stronger winds found in models with larger vertical magnetic flux and higher spin.  Both strong and weak jets are produced: strong jets are persistent, highly relativistic, and magnetically driven, while weak jets are intermittent, mildly relativistic, and powered by a combination of magnetic and radiative forces. 
\end{abstract}

\keywords{\uat{Radiative magnetohydrodynamics}{2009} --- \uat{General relativity}{641} --- \uat{Black hole physics}{159} --- \uat{Accretion}{14} --- \uat{Ultraluminous X-ray sources}{2164} --- \uat{X-ray binary stars}{1811}}

\section{Introduction}
\label{sec:introduction}

The Eddington limit is conventionally defined by the balance between gravity and radiation force under the assumption of spherical symmetry, and thus sets a characteristic scale for luminosity and accretion rate.  In realistic accretion systems, however, this limit is not a strict barrier: geometric collimation \citep[e.g.,][]{King2009}, photon trapping \citep[e.g.,][]{Begelman1978}, anisotropic radiation beaming \citep[e.g.,][]{Middleton2015}, and magnetic support \citep[e.g.,][]{Begelman2007} can all permit accretion to exceed the classical Eddington limit.  The near-Eddington regime is therefore particularly intriguing, as it captures the regime transition between sub- and super-Eddington flows.  In this regime, radiation in the inner disk becomes highly advective due to photon trapping, resembling slim disk behavior \citep{Abramowicz1988}, while the outer disk remains largely diffusive, which is consistent with Novikov-Thorne or $\alpha$-disk assumptions \citep{Novikov1973,Shakura1973}. 

Observationally, near-Eddington accretion is relevant to luminous X-ray binaries \citep{McClintock2006,Done2007}, ultraluminous X-ray sources \citep{Kaaret2017,King2023}, and rapidly growing black holes (e.g., those powering high-redshift quasars and massive black hole seeds; \citealt{Inayoshi2020}).  In these systems, spectral and timing properties often deviate from both standard thin disk predictions and fully supercritical expectations \citep[e.g.,][]{Poutanen2007,Gladstone2009,Middleton2015}.  Moreover, because of the intrinsically anisotropic structure of near-Eddington flows, the apparent luminosity can be strongly viewing-angle dependent, spanning roughly $0.1$-$10L_{\mathrm{Edd}}$ (see \autoref{sec:observational_implications} for discussion).  As a result, near-Eddington accretion systems are difficult to identify observationally based on luminosity alone, highlighting the need for improved theoretical diagnostics.

Despite its importance, the near-Eddington regime remains comparatively underexplored in theory.  Unlike the sub- or super-Eddington regimes, it does not admit simplifying assumptions that decouple radiation, turbulence, or magnetic fields.  Instead, it requires the simultaneous treatment of radiation transport, MRI-driven turbulence, and global magnetic structure.  

Several numerical studies have addressed this regime.  \citet{Sadowski2016} presented GRMHD simulations with M1 radiation transport, showing that magnetic field topology plays an important role in dynamics.  Follow-up work by \citet{Lancova2019} and \citet{Wielgus2022} focused on dipolar magnetic configurations and identified a steady ``puffy disk'' solution, consistent with earlier isothermal models reported by \citet{Zhu2018} and \citet{Mishra2020}.  More recently, \citet{Fragile2023,Fragile2025} carried out a series of GRMHD simulations with M1 radiation transport, with the former emphasizing multi-frequency radiation effects and the latter exploring a broad set of initial conditions.  Meanwhile, \citet{Huang2023} performed a parameter study of magnetic topology and accretion rate using Newtonian MHD coupled with full radiation transport.  

Nevertheless, the relevant parameter space remains incompletely explored.  \citet{Sadowski2016} and \citet{Lancova2019} considered only non-spinning black holes, and some models were under- or marginally resolved due to computational limitations.  \citet{Fragile2023} and \citet{Fragile2025} focused exclusively on the quadrupolar magnetic topology, while the Newtonian framework adopted by \citet{Huang2023} cannot capture relativistic dynamics in the vicinity of the black hole.  Detailed comparisons with these previous studies are presented in \autoref{sec:compare_other_models}.  

Recently, we have carried out a series of radiation GRMHD simulations to investigate radiation-dominated accretion flows onto black holes using a full radiation transport algorithm \citep{White2023}.  These simulations are performed with the GPU-accelerated code \texttt{AthenaK} \citep{Stone2024}, enabling the high-resolution calculations required on modern exascale computing systems.  The simulation suite spans a broad parameter space, covering a wide range of accretion rates, two black hole spins, and two magnetic field topologies.  In our first paper (\citealt{PaperI}, hereafter \citetalias{PaperI}), we present an overview of the simulation suite and its global properties.  The second paper (\citealt{PaperII}, hereafter \citetalias{PaperII}) focuses on the super-Eddington regime, providing a detailed analysis and establishing a unified analysis framework for the entire series.  Subsequent papers therefore adopt this analysis routine without repetition and refer readers to \citetalias{PaperII} for methodological details. 

This is the third paper in the series, which focuses on near-Eddington accretion models and provides a detailed analysis of steady-state solutions across different magnetic field topologies, as well as a systematic discussion of general relativistic effects, including black hole spin and jet formation.  In the next paper (Paper IV), we will further present and analyze the sub-Eddington accretion models.

\section{Numerical Methods}
\label{sec:numerical_methods}

We perform the simulations by solving the 3D radiation GRMHD equations in Cartesian Kerr-Schild coordinates.  The angle-dependent radiation intensity is evolved directly through the radiative transfer equation in the local tetrad frame \citep{White2023}, with the radiation source terms described in equation~(4) of \citetalias{PaperI}.  Details of the simulation setup, including the mesh configuration and the initial and boundary conditions, are provided in Section~2 of \citetalias{PaperI} and are not repeated here. 

Unless otherwise specified, variables are defined in the coordinate frame.  Overbars denote fluid-frame quantities, and tildes indicate power-law fitted values. Additional notation details can be found in \citetalias{PaperI}. 

We adopt an intermediate-resolution mesh for all near-Eddington models except E09-a3-DL, which develops a magnetically elevated disk that is already well resolved on the low-resolution grid.  Each intermediate-resolution model also has a corresponding low-resolution version for the resolution study.  

Both the intermediate- and low-resolution grids achieve the same maximum spatial resolution near the black hole, with a minimum cell size of $0.0625r_g$.  However, the intermediate-resolution mesh covers an eight-times larger volume in the most refined region and provides twice the spatial resolution over most of the computational domain compared to the low-resolution mesh (see Section~2.3 of \citetalias{PaperI}).  The number of cells crossing the origin in the $x$, $y$, and $z$ directions are $2048\times2048\times1152$ at intermediate resolution, approximately twice those of the low-resolution grid in each direction ($1152\times1152\times640$). 

\begin{deluxetable*}{l c c c c c c c c}
\tablecaption{Comparison of Time- and Azimuthally Averaged Properties of the Near-Eddington Models \label{tab:sim_overall}}
\tablehead{
\colhead{Name} & 
\colhead{Type} & 
\colhead{$\rho_0$} & 
\colhead{$\Delta z_{\mathrm{disk},20}$} & 
\colhead{$r_{\mathrm{eq}}$} & 
\colhead{$\tilde{P}_{\mathrm{therm}}^{(\mathrm{disk})}$} & 
\colhead{$\tilde{P}_{\mathrm{m}}^{(\mathrm{disk})}$} & 
\colhead{$\tilde{\alpha}_{\mathrm{Max}}^{r~(\mathrm{disk})}$} & 
\colhead{$\tilde{\alpha}_{\mathrm{Rey,turb}}^{r~(\mathrm{disk})}$}
\\
& &
\colhead{$\left(\mathrm{g/cm^3}\right)$} &  
\colhead{$(r_g)$} & \colhead{$(r_g)$} & 
\colhead{$(10^{13}~\mathrm{dyn/cm^{2}})$} & \colhead{$(10^{13}~\mathrm{dyn/cm^{2}})$} & 
\colhead{$(10^{-2})$} & \colhead{$(10^{-2})$}
\\
\quad\;\;\;(1) & (2) & (3) & (4) & (5) & (6) & (7) & (8) & (9)
}
\startdata
    E09-a3-DL & Mag & 3e-5 & 36 & 60 & $2.89~r_{10}^{-1.32}$ & $5.42~r_{10}^{-1.30}$ &    $14.40~r_{10}^{-0.71}$ & $2.45~r_{10}^{-0.12}$ \\ 
    E08-a9 & Therm    & 2e-3 & 29 & 30 & $4.66~r_{10}^{-1.73}$ & $4.39~r_{10}^{-1.18}$ &    $11.75~r_{10}^{-0.44}$ & $2.27~r_{10}^{-0.10}$ \\ 
    E08-a3 & Therm    & 8e-4 & 31 & 28 & $2.24~r_{10}^{-1.48}$ & $3.69~r_{10}^{-1.35}$ &    $20.35~r_{10}^{-0.94}$ & $2.33~r_{10}^{-0.19}$ \\ 
    E07-a3-DL & Therm & 8e-5 & 20 & 21 & $6.11~r_{10}^{-0.71}$ & $3.40~r_{10}^{-0.86}$ & \;\:$7.99~r_{10}^{-0.21}$ & $1.81~r_{10}^{-0.48}$ \\ 
    \hline
\enddata
\tablecomments{
    {\bf Columns (from left to right):} 
    (1) Model name (`E': accretion rate in Eddington units; `a': black hole spin); 
    (2) Disk type; 
    (3) Simulation density unit; 
    (4) Disk vertical thickness at $r=20r_g$; 
    (5) Maximum inflow equilibrium radius; 
    (6) Power-law fitted, disk-averaged thermal pressure (gas + radiation); 
    (7) Power-law fitted, disk-averaged magnetic pressure; 
    (8) Power-law fitted, disk-averaged radial Maxwell component of the angular momentum flux, normalized by the total pressure (equation~13 in \citetalias{PaperII}); 
    (9) Power-law fitted, disk-averaged radial turbulent Reynolds component of the angular momentum flux, normalized by the total pressure (equation~13 in \citetalias{PaperII}). 
    More model parameters and measurements are provided in Table 1 of \citetalias{PaperI}.
}
\end{deluxetable*}

In total, we run 7 simulations, 4 of which are summarized in \autoref{tab:sim_overall} and used for the main results, while the remaining 3 are used for resolution comparison.  The model naming convention is described in Section~2.4 of \citetalias{PaperI}.  Because the radiative models are not scale-free, we set the density unit $\rho_0$ to specify the accretion rate; the value of $\rho_0$ adopted for each model is listed in \autoref{tab:sim_overall}.  All other details related to the numerical setup and relevant definitions can be found in Section~2 of both \citetalias{PaperI} and \citetalias{PaperII}.  

\section{Results} 
\label{sec:results}

In this section, we present four near-Eddington accretion models that differ in black hole spin ($a=0.3$ and $a=0.9375$) and in their initial magnetic field topology (single-loop and double-loop configurations).  The models consist of two single-loop cases with different spins and two double-loop cases with different initial accretion rates.  These simulations result in two distinct steady-state solutions (e.g., see Figure~8 in \citetalias{PaperI}): (1) a thin thermal disk embedded within a magnetically dominated envelope, and (2) a magnetically elevated disk.  

Among the four models, only the one initialized with a double-loop magnetic configuration develops a magnetically elevated disk, while the remaining three, including another double-loop run starting at a higher accretion rate, remain thin thermal disks in the final steady state.  The thin thermal disk are characterized by the presence of a net vertical magnetic flux through the disk midplane, whereas the magnetically elevated disk shows weak or no net vertical flux. 

The single- and double-loop setups were originally designed to yield, respectively, the maximum and zero net vertical flux for comparison.  Interestingly, the double-loop model initialized with a larger density scale, and therefore a higher initial accretion rate, also develops a net vertical flux.  This arises because strong, anisotropic, radiation-driven outflows disrupt the initial magnetic polarity, allowing vertical flux to accumulate in the inner disk and driving the system toward a thin thermal disk similar to the single-loop cases.  The detailed discussion of this evolution is provided in \autoref{sec:magnetic_polarity_breaking}.  

We summarize the final steady-state disk type for each model in \autoref{tab:sim_overall}, where `Mag' denotes a magnetically elevated disk and 'Therm' denotes a thin thermal disk.  In the following subsections, we present the simulation results and analyze the structures and physical properties of the systems, emphasizing the differences between the two characteristic disk solutions.  Details of the analysis methods are provided in \citetalias{PaperII}.

\subsection{Time Evolution}

\begin{figure*}
    \centering
    \includegraphics[width=\textwidth]{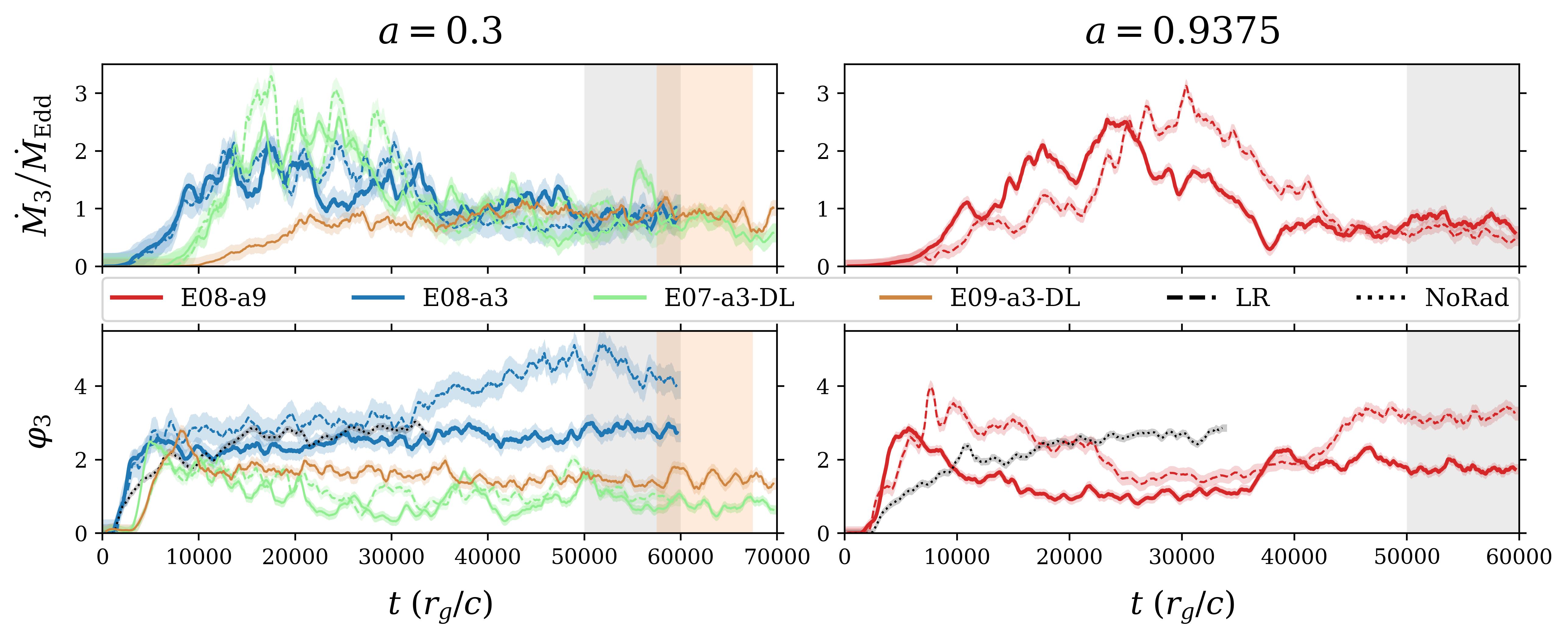}
    \caption{
    Time evolution of the mass accretion rate $\dot{M}_3$ (top) and normalized magnetic flux $\varphi_3$ (bottom) measured at $r=3r_g$; the subscript `3' denotes quantities evaluated at this radius.  The accretion rate is expressed in units of the Eddington accretion rate, assuming a 10\% radiative efficiency.  The definitions of all quantities are provided in Section~3.1 of \citetalias{PaperI}.  The left panels show models with low black hole spin, while the right panels correspond to high-spin cases.  Each time series is smoothed using a moving boxcar with a time window of $800r_g/c$ and shaded in the same color to indicate the $1\sigma$ variation within the local time window.  Line colors represent different accretion rates, as indicated in the legend.  Low-resolution counterparts, where applicable, are shown by dashed lines.  The black dotted lines in the lower panels denote the non-radiative single-loop counterparts (see NoRad-a3 and NoRad-a9 in \citetalias{PaperII}).  Color-shaded time intervals mark the epochs used for time-averaged analysis of the steady-state structure, with orange for model E07-a3-DL and gray for the others. 
    }
    \label{fig:hst_compare}
\end{figure*}

\begin{figure*}
    \centering
    \includegraphics[width=\textwidth]{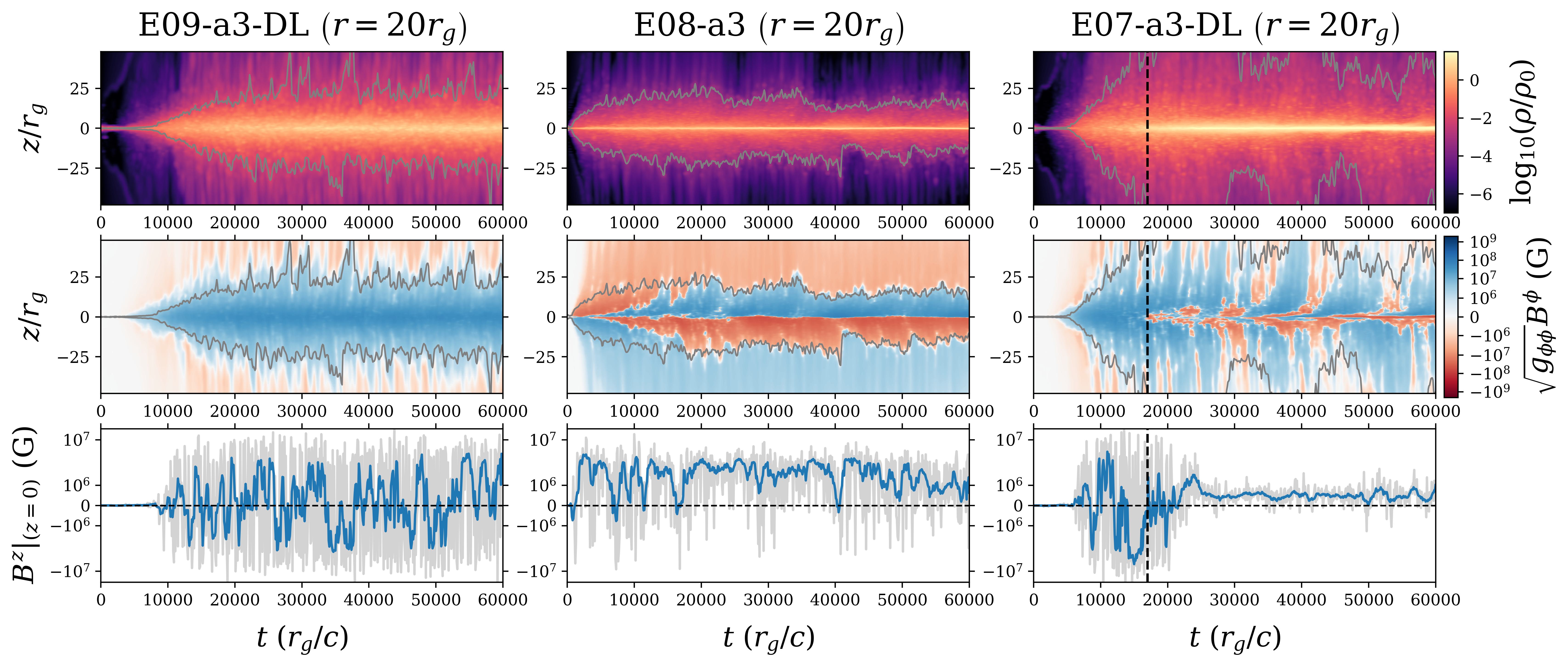}
    \caption{ 
    Time evolution of gas density, toroidal magnetic field, and midplane vertical magnetic field measured at a cylindrical radius of $20r_g$.  In the space-time diagrams of gas density (top) and toroidal magnetic field (middle), gray dashed lines mark the scattering photosphere.  In the time series of the midplane vertical magnetic field (bottom), gray lines show the raw data sampled at $\Delta t=10r_g/c$, and blue lines show the moving-boxcar average with a time window of $100r_g/c$.  The vertical dashed lines in E07-a3-DL marks the time ($t=17000r_g/c$) when the system evolves from a magnetically elevated flow into a steady thin thermal disk as vertical magnetic flux accumulates.
    } 
    \label{fig:timestream}
\end{figure*}

As shown in \autoref{fig:hst_compare}, all models reach steady states in both accretion rate ($\dot{M}_3$) and magnetic flux ($\varphi_3$) measured near the black hole at $3r_g$ after $t=40000r_g/c$, remaining in the near-Eddington regime and the SANE state with $\varphi_3 \lesssim 5$.  For the thin thermal disk models (E08-a3, E08-a9, and E07-a3-DL), the accretion rate initially rises slightly above the Eddington limit, then gradually declines and settles around $0.8\dot{M}_{\mathrm{Edd}}$.  In contrast, the magnetically elevated disk (E09-a3-DL) maintains a nearly constant accretion rate close to the Eddington level till the end of the simulation.  

The magnetic flux stabilizes earlier than the accretion rate in all cases. It shows no clear dependence on disk type, although the double-loop runs generally exhibit smaller fluxes than the single-loop runs.  This trend is consistent with the super-Eddington models (see Figure~2 in \citetalias{PaperII}), suggesting that the steady state of magnetic flux is largely independent of the accretion regime.  However, this does not necessarily imply the same dynamical behavior.

\autoref{fig:timestream} shows the time evolution of gas density (top row) and magnetic fields (middle and bottom rows) at a cylindrical radius of $20r_g$.  The vertical magnetic field, shown in the bottom row, is measured at the midplane.  These near-Eddington models demonstrate a strong dependence of disk morphology on the presence of vertical magnetic flux.    

When the system has net vertical flux through the midplane (E08-a3 and the late phase of E07-a3-DL, with vertical field $\gtrsim 5\times10^5$~G at $20r_g$), it develops a density concentration near the midplane.  In contrast, accretion flow without net vertical flux (E09-a3-DL and the early phase of E07-a3-DL) forms magnetically elevated disks lacking a dense midplane layer.  Note that midplane of model E07-a3-DL is initially magnetically elevated but later collapses into a dense layer as net vertical flux accumulates, marked by the vertical dashed lines in the right column of \autoref{fig:timestream}.

The space-time diagrams of the toroidal magnetic field (middle row of \autoref{fig:timestream}) in the near-Eddington regime depend strongly on the strength of the net vertical magnetic flux: (1) in magnetically elevated disks without net vertical flux (E09-a3-DL and the early phase of E07-a3-DL), the toroidal field maintains a single polarity across the midplane; (2) with strong net vertical flux (E08-a3), it shows opposite polarities above and below the midplane; and (3) with weak net vertical flux (late phase of E07-a3-DL), it exhibits field-reversal patterns near the midplane.  

Unlike the super-Eddington models, which all display butterfly patterns (see Figure~3 in \citetalias{PaperII}), only case (3) shows magnetic polarity reversals in the near-Eddington regime, whereas cases (1) and (2) maintain constant polarity consistent with the shear of the global magnetic field, indicating dominance of the mean magnetic field.

\subsection{Steady State}

\begin{figure}
    \centering
    \includegraphics[width=\columnwidth]{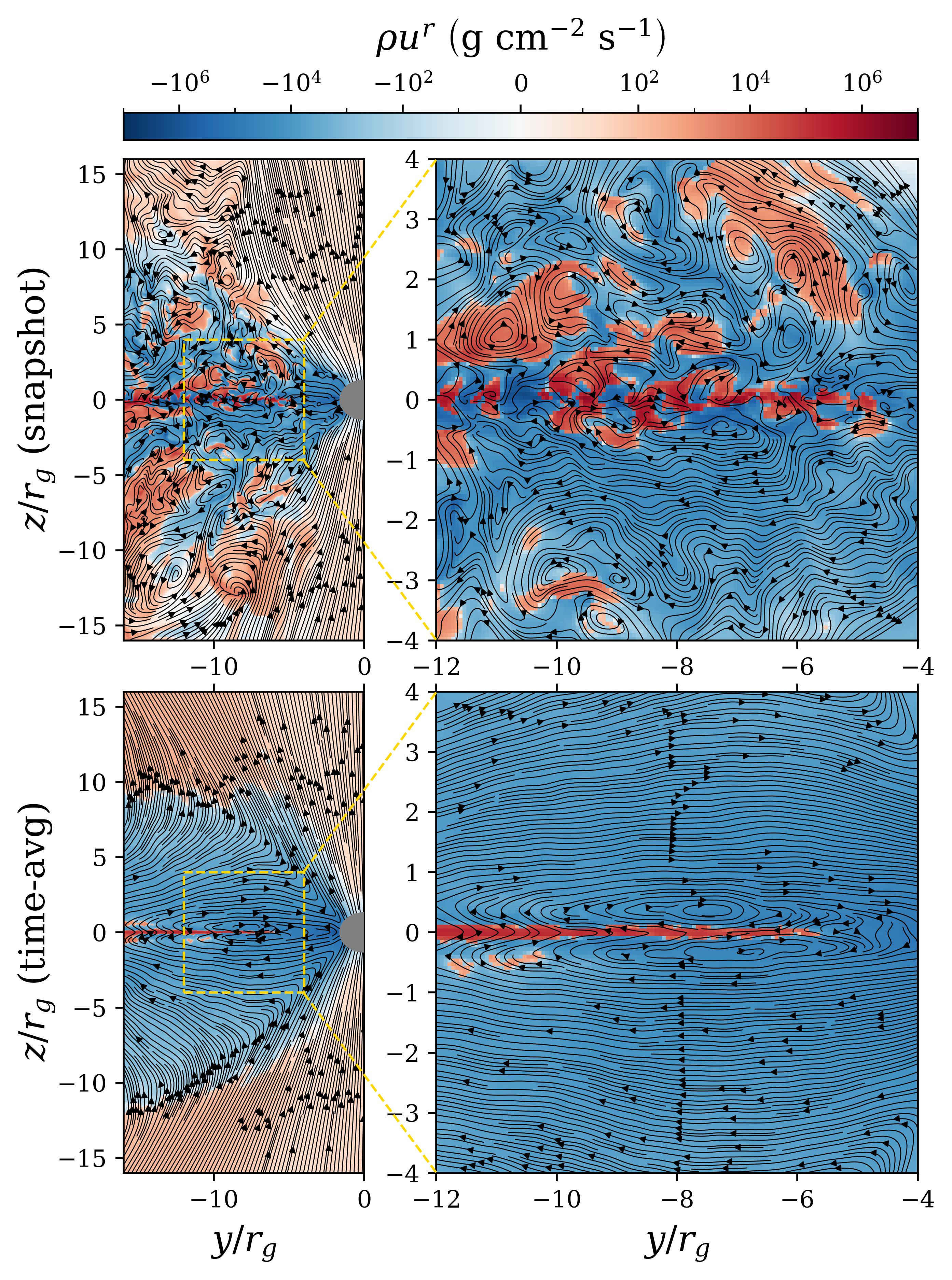}
    \caption{
    Comparison of radial gas density flow (colormap) and magnetic field lines (streamlines) between an instantaneous snapshot (top) and the time-averaged structure (bottom) in the steady state, using model E08-a9 as an example.  The left panels show the global view, while the right panels present the zoomed-in regions outlined by yellow dashed boxes. Red indicates gas moving radially outward and blue indicates gas flowing inward.  Both the gas flow and magnetic field are highly turbulent in the snapshot but well organized in the time average.  Other near-Eddington models exhibit similar behavior.  
    }
    \label{fig:turb_compare}
\end{figure}

\begin{figure*}
    \centering
    \includegraphics[width=\textwidth]{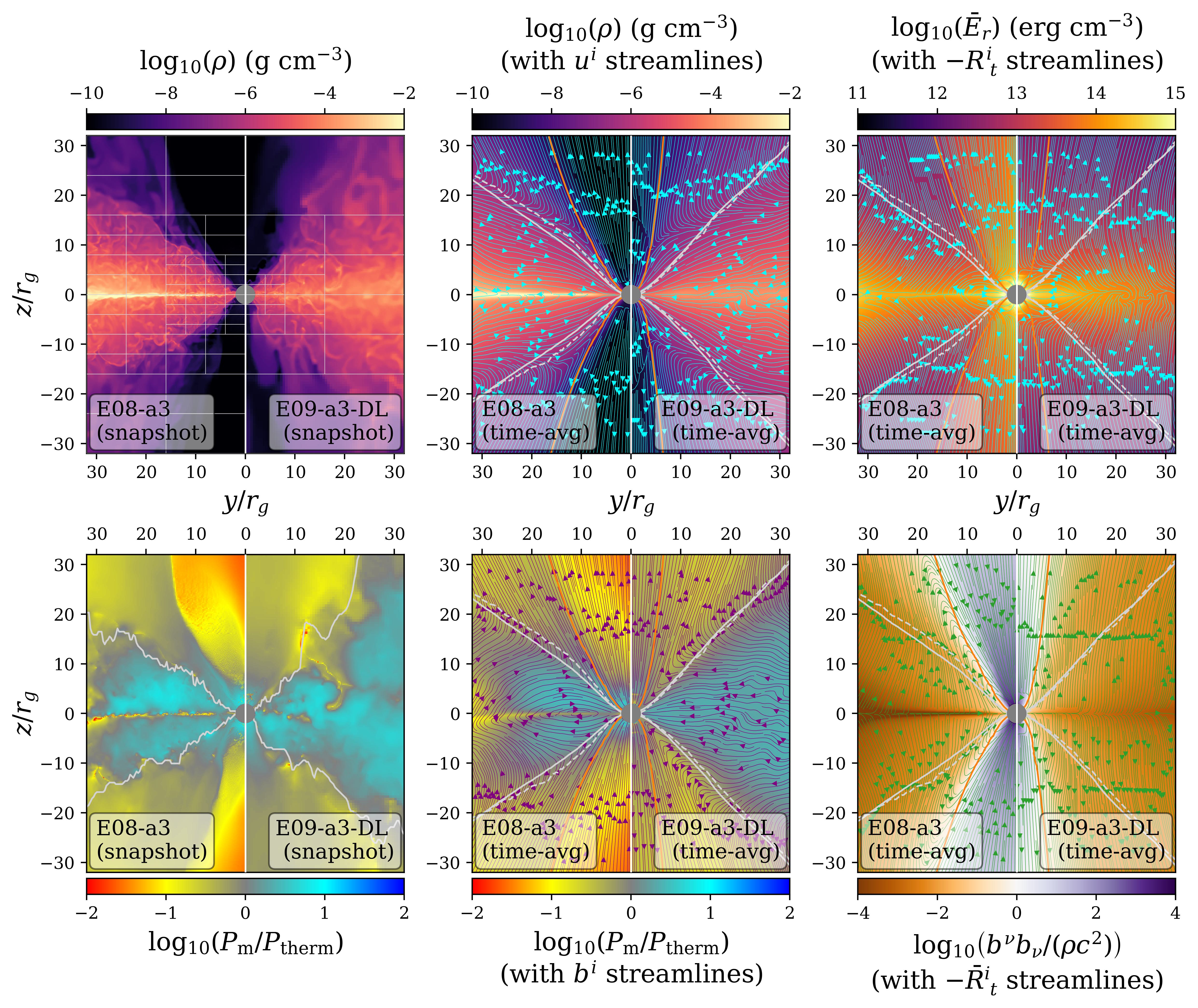}
    \caption{
    2D profiles of the two stead-steady solutions in the near-Eddington regime, showing gas density (upper left and middle), fluid-frame radiation energy density (upper right), magnetic-to-thermal pressure ratio (lower left and middle), and magnetization (lower right).  The thermal pressure includes both gas and radiation contributions.  Each panel presents a side-by-side comparison between the thin thermal disk (left half) and the magnetically elevated disk (right half), using models E08-a3 and E09-a3-DL as examples, respectively.  The left column shows instantaneous snapshots at $t=60000r_g/c$, while the middle and right columns show the time-averaged profiles over $t=50000$-$60000r_g/c$.  Different streamlines are overlaid on the time-averaged panels: velocity in the density panels, magnetic field in the pressure-ratio panels, coordinate-frame radiation flux in the radiation-energy panel, and fluid-frame radiation flux in the magnetization panel.  The mesh block configuration is outlined with gray solid lines in the density snapshot.  Auxiliary lines help distinguish different regions: orange lines mark jet boundaries, solid gray lines trace the scattering photosphere, and gray dashed lines denote disk boundaries defined by zero Bernoulli parameter.  All related definitions are provided in Section~3.2 of \citetalias{PaperII}. 
    }
    \label{fig:profile2d}
\end{figure*}

After the systems reach a steady state, the accretion flow remains turbulent, but its time-averaged structure is remarkably coherent, as shown in \autoref{fig:turb_compare}.  In the instantaneous snapshot (top), local turbulent features, such as magnetic field loops, closely follow the gas flow, but these features vanish in the time-averaged view.  This indicates that the system is dominated primarily by the mean magnetic field and mean flow rather than by turbulence.  Such organization implies that both the dynamics and flow structure depend sensitively on the global magnetic field topology, consistent with these two magnetically dependent steady-state solutions found in the near-Eddington regime.  In contrast, the super-Eddington systems are turbulence-dominated and exhibit only weak dependence of their flow structure on the field topology (see \citetalias{PaperII} for details). 

We identify the region that reaches inflow equilibrium (as described in Section~3.2 of \citetalias{PaperII}) and report its outermost radius for each model in \autoref{tab:sim_overall}.  We also measure the disk vertical extent at $20r_g$ within the inflow equilibrium region for all near-Eddington models and summarize these values in \autoref{tab:sim_overall}. 

\autoref{fig:profile2d} compares the two steady-state solutions, with each panel showing a side-by-side comparison of the thin thermal disk (left half) and the magnetically elevated disk (right half).  Unlike the magnetically elevated disk, which is dominated by magnetic pressure throughout, the thin thermal disk is embedded within an optically thick magnetic envelope, with gas density strongly concentrated near the midplane.  Accretion in the inner region of the thin disk is peaked around the midplane; however, at larger radii ($\gtrsim 10r_g$), the midplane gas moves outward, and accretion occurs primarily in the magnetic envelope above the midplane.  This surface-dominated accretion closely resembles that reported by \citet{Zhu2018} and differs from the magnetically elevated disk, where accretion proceeds throughout the entire disk body.  

The global magnetic field structures differ markedly between the two disk solutions.  The magnetically elevated disk retains the initial double-loop topology, featuring oppositely directed poloidal magnetic fields above and below the midplane.  In contrast, the thin thermal disk develops torus-like magnetic loops centered around the midplane within the disk body; outside the disk, the field lines open outward and preserve the initial single-loop polarity.  The magnetic loops inside the disk are oriented opposite to the initial field direction (see Figure~1 in \citetalias{PaperI}), indicating that this configuration arises not from a simple compression of the initial field, but from the magnetic-envelope accretion: the inflow moves faster at higher altitudes, carrying magnetic field lines inward and pinching the field along the disk surface.

By comparing the radiation fluxes in the coordinate and fluid frames, we can quantify the relative role of radiation advection and diffusion in heat transport.  In the super-Eddington regime, radiation is strongly advective (see Figure~5 in \citetalias{PaperII}), particularly near the midplane where it is largely trapped and carried inward by the accretion flow.  In contrast, in the near-Eddington regime, radiation advection has only a moderate influence: it remains somewhat advective within $\sim10r_g$, but becomes highly diffusive at larger radii, where radiation escapes predominantly perpendicular to the disk.  These behaviors are also reflected in Figure~5 of \citetalias{PaperI}: in the super-Eddington models, the thermal time (including both radiation diffusion and advection) is much shorter than the diffusion time, indicating advection dominance, whereas in near-Eddington models, the thermal and diffusion times are comparable, suggesting diffusion dominance. 

The wind region lies above the disk (gray dashed lines) and below the jet (orange solid lines).  The disk region is defined by the contour of zero Bernoulli parameter, while the jet boundary is traced by velocity streamlines originating from the strongly magnetized funnel.  Detailed definitions of the system partition are provided in Section~3.2 of \citetalias{PaperII}.  The wind is primarily thermal and launched from the disk surface, but the outflows are mostly optically thin, unlike in the super-Eddington models, where the wind can be mostly optically thick.  Jets form in the strongly magnetized funnel region: single-loop models produce powerful relativistic jets in which radiation exerts a drag on the plasma, whereas double-loop models can produce intermittent, mildly relativistic jets where radiation instead exerts a push, as indicated by the direction of the fluid-frame radiation flux.  

\subsection{Disk Properties}

\subsubsection{Radial Structure}
\begin{figure}
    \centering
    \includegraphics[width=\columnwidth]{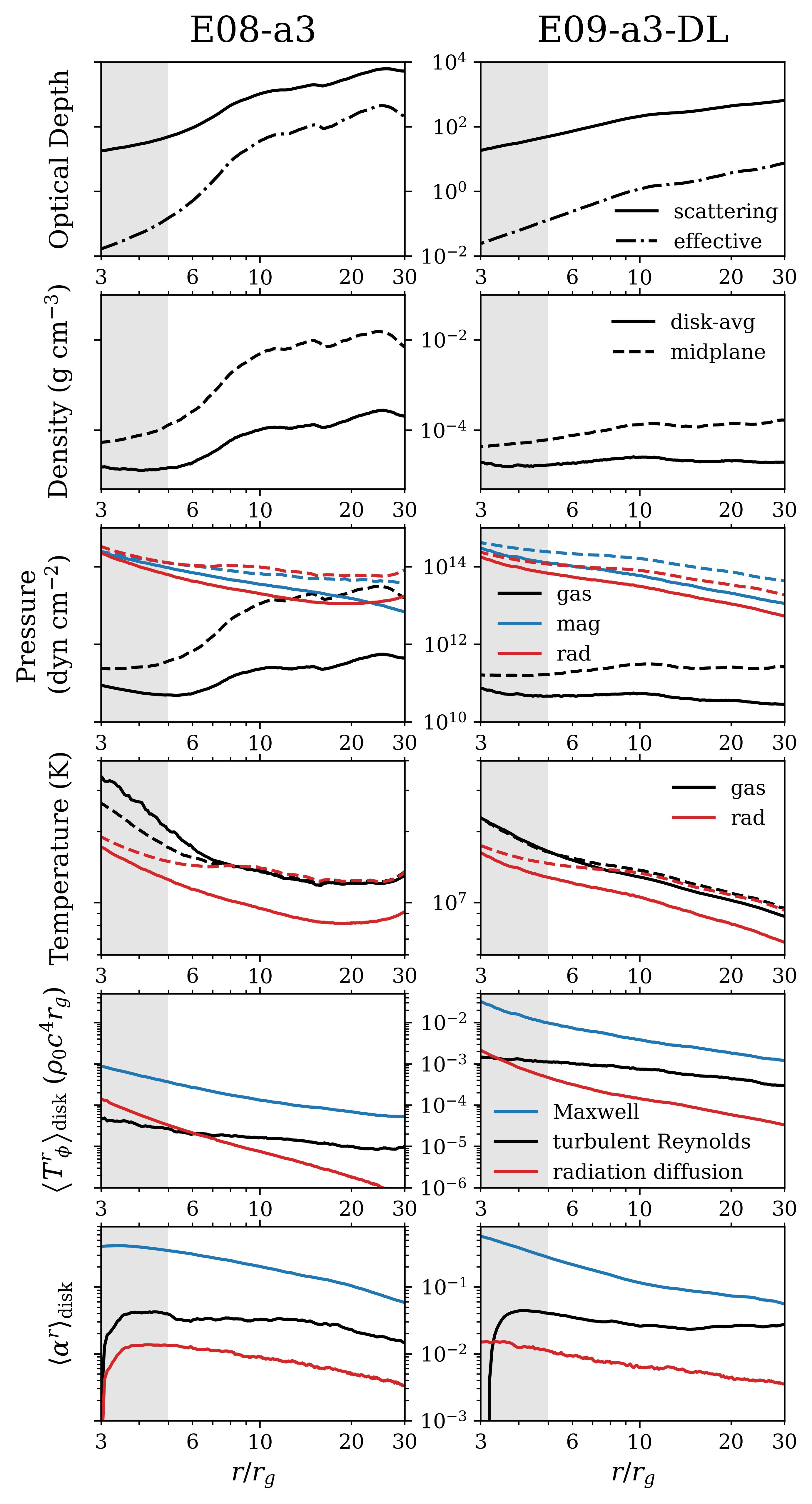}
    \caption{
    Radial profiles of two representative near-Eddington models in time- and azimuthal average.  The left column corresponds to the thin thermal disk embedded within a magnetic envelope, and the right column to the magnetically elevated disk.  The gray shaded region marks the plunging region.  The first row shows the disk optical depths: scattering (solid) and effective (dot-dashed).  The remaining rows present density, pressure, temperature, radial angular momentum flux, and effective viscosity, each measured as disk averages (solid) or at the midplane (dashed, where applicable).  Gas, magnetic, and radiation components in the pressure and temperature panels are shown in black, blue, and red, respectively.  In the angular momentum flux and effective viscosity panels, turbulent Reynolds, Maxwell, and radiation diffusion stresses are likewise shown in black, blue, and red.  Details on how these quantities are computed are provided in Section~3.3 and 3.5 of \citetalias{PaperII}.  
    }
    \label{fig:hori_compare}
\end{figure}

\autoref{fig:hori_compare} presents the radial profiles of the two steady-state disk solutions in the near-Eddington regime (i.e., model E08-a3 representing the thin thermal disk within a magnetic envelope, and model E09-a3-DL representing the magnetically elevated disk), showing the optical depth, density, pressure, temperature, angular momentum flux, and effective viscosity.  The definitions of all quantities are provided in Section~3.3 and 3.5 of \citetalias{PaperII}.  

Both the scattering and effective optical depths of the disk body increases with radius, with the scattering optical depth generally larger because the Planck-mean opacity is much lower than the Thomson opacity in this high-temperature regime.  In the inner disk region ($\lesssim 8r_g$), the effective optical depth falls below unity, indicating that the gas and radiation are not sufficiently coupled to maintain thermal equilibrium. 

The density contrast between the midplane (dashed) and the disk average (solid) is much larger in the thin thermal disk -- by up to about two orders of magnitude -- than in the magnetically elevated disk, reflecting the strong midplane density concentration in the former.  The density increases radially in the thin disk but remains nearly flat in the magnetically elevated disk.  The overall density dependence arises from the requirements of force (momentum) and thermal (energy) equilibrium.  In magnetically or radiation-supported systems, gas density contributes little to the force balance and enters primarily through radiation-gas thermal coupling, leading to only a weak density dependence.  In contrast, the thin thermal disk shows a strong radial density dependence because its midplane is strongly supported by gas pressure force through density concentration (see \autoref{sec:vertical_structures} for details), unlike the magnetically elevated disk, which is supported by magnetic and radiation forces. 

Radiation and magnetic pressures dominate over the gas pressure in both steady-state disk solutions.  However, in the thin thermal disk, the midplane gas pressure can rise to comparable levels (but still less) owing to the strong density concentration.  Gas and radiation remain in thermal equilibrium near the midplane, except in the inner disk region where their temperatures begin to diverge, consistent with the low effective optical depth.  The gas temperature is generally higher than the radiation temperature in the disk average because the gas becomes thermally decoupled from the radiation field near the disk surface. 

\begin{figure}
    \centering
    \includegraphics[width=\columnwidth]{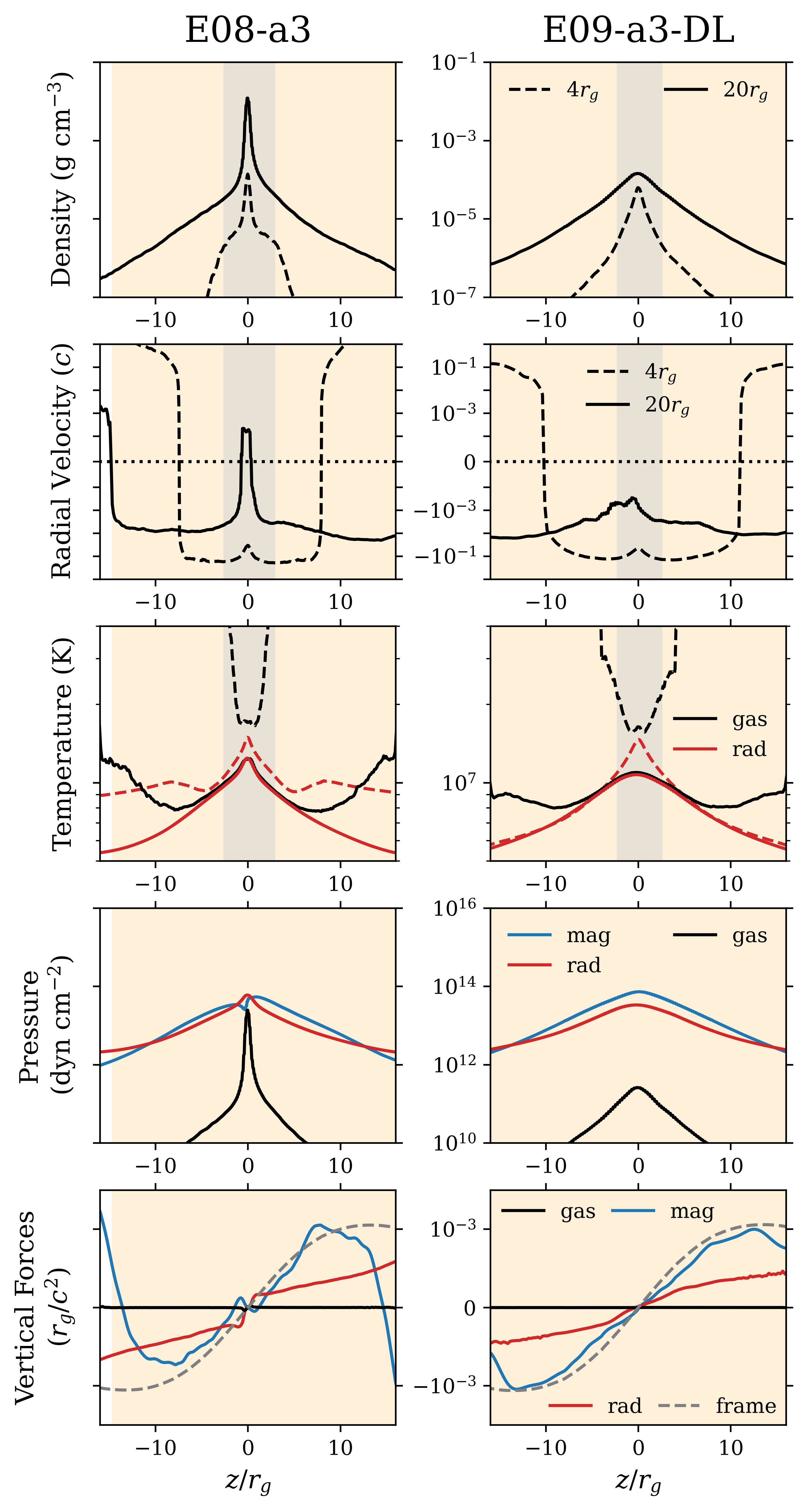}
    \caption{
    Vertical profiles of two representative near-Eddington models in time- and azimuthal average.  The left column corresponds to the thin thermal disk embedded within a magnetic envelope, and the right column to the magnetically elevated disk.  From top to bottom, the rows show density, radial velocity, temperature, pressure, and vertical forces, measured at horizontal distances of $20r_g$ (solid) or $4r_g$ (dashed, where applicable), with the corresponding gravitationally bounded disk region highlighted in yellow or gray, respectively.  Gas, magnetic, and radiation components are plotted in black, blue, and red, respectively.  The horizontal dotted line in the velocity panel marks zero velocity.  In the bottom row, the frame force (dominated by gravity) is plotted with its sign inverted ($-f_{\mathrm{frame}}$) as gray dot-dashed lines for comparison.  Details on the computation of four-forces are provided in Section~3.4 of \citetalias{PaperII}. 
    }
    \label{fig:vert_compare}
\end{figure}

Both the disk-averaged angular momentum flux and the corresponding effective viscosity show that the accretion process is driven by outward angular momentum transport primarily due to the Maxwell stress.  The turbulent Reynolds stress is subdominant, while the contribution of radiation is nearly negligible.  However, near the horizon, as the radiation field becomes thermally decoupled from the gas, it also becomes increasingly anisotropic.  This enhances the off-diagonal components of the radiation stress tensor, allowing the radiation stress to locally exceed the Reynolds stress.  Nevertheless, because the Maxwell stress remains dominant, this local effect has little impact on the overall disk dynamics.

\begin{figure*}
    \centering
    \includegraphics[width=\textwidth]{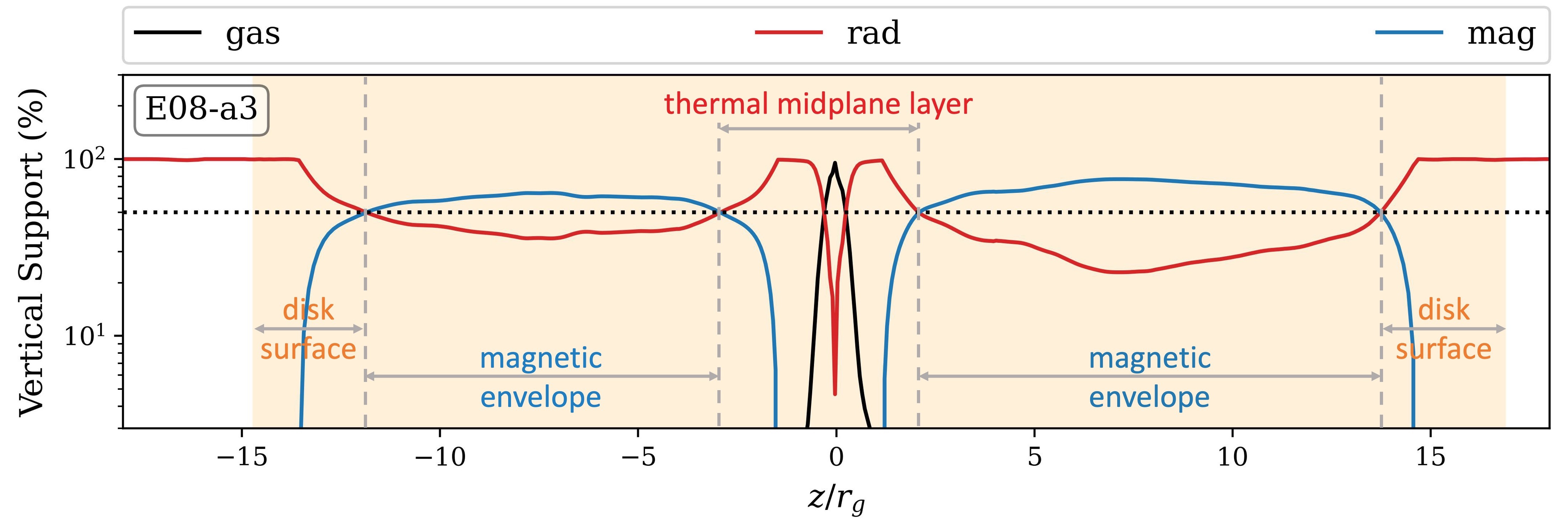}
    \caption{    
    Percentage contributions to the vertical support from gas (black), magnetic fields (blue), and radiation (red) in the thin thermal disk model E08-a3, measured at a horizontal distance of $20r_g$ using time- and azimuthal averages.  For each four-force component that provides support against gravity, we compute its contribution as a percentage of the total vertical support.  The definitions of the four-force terms ($f_{\mathrm{gas}}$ for gas pressure force, $f_{\mathrm{pmag}}+f_{\mathrm{tmag}}$ for the total magnetic force, and $f_{\mathrm{rad}}$ for the radiation force) are given in Section~3.4 of \citetalias{PaperII}. 
    }
    \label{fig:vert_support}
\end{figure*}

The radial profiles of pressure, temperature, angular momentum flux, and effective viscosity generally follow power laws.  We summarize the fitting parameters for pressure and effective viscosity in \autoref{tab:sim_overall}, while the disk temperature are provided separately in Table~1 of \citetalias{PaperI}.  For model E07-a3-DL, although we provide an overall fit, the Maxwell component of the effective viscosity shows two distinct power-law segments, with a break near $8r_g$.  To be more specific, we fit the two regions separately, obtaining $11.69\times10^{-2} r_5^{-0.92}$ for the inner part and $7.59\times10^{-2} r_{10}^{-0.06}$ for the outer part, where the subscripts indicate the radius at which the quantities are measured.  This distinct behavior is likely related to the magnetic field evolution during the disk transition (see discussions in \autoref{sec:angular_momentum_transport} and \autoref{sec:magnetic_polarity_breaking} for more details). 

\subsubsection{Vertical Structure}
\label{sec:vertical_structures}

\autoref{fig:vert_compare} presents the vertical structures, with the relevant definitions of all quantities provided in Section~3.4 of \citetalias{PaperII}.  Both the thin thermal disk and the magnetically elevated disk exhibit density concentration toward the midplane, although the thin disk shows a much stronger concentration.  

The two disks exhibit distinct accretion patterns.  The thin thermal disk accretes primarily through inflow in the magnetic envelope, with much weaker accretion or even decretion near the midplane, whereas the magnetically elevated disk accretes more uniformly throughout its vertical extent. 

Except at the innermost radii, gas and radiation remain thermally coupled near the midplane but gradually decouple toward the disk surface, forming a thermally decoupled layer that spans roughly the upper half of the disk.  

Throughout the vertical extent, the pressure is dominated by either radiation or magnetic components, while radiation pressure consistently dominates near the disk surface. 

The two disks also differ fundamentally in their vertical support structures.  In the thin thermal disk, gas pressure increases sharply toward the midplane due to strong density concentration; although its magnitude remains below that of the radiation and magnetic pressures, its steep time-averaged vertical gradient provides the primary support against gravity near the midplane.  Magnetic pressure peaks above the midplane and therefore compresses rather than supports the midplane layer, while radiation pressure contributes across the full vertical extent and becomes dominant toward the disk surface.  By contrast, the magnetically elevated disk is supported primarily by magnetic forces throughout most of its vertical extent. 

\autoref{fig:vert_support} shows the percentage contribution to the vertical support from each component (gas, magnetic, and radiation) in the thin thermal disk model.  The thin thermal disk itself is supported only by gas and radiation, with gas pressure force dominating near the midplane.  In the overlying magnetic envelope, magnetic forces provide the primary support with radiation subdominant, until near the disk surface where radiation overtakes magnetic forces and becomes dominant.  

Note that this vertical support analysis is based on time-averaged quantities.  In instantaneous snapshots of the thin thermal disk, the high-density region is highly oscillatory and closely follows the global current sheet that forms in the disk.  However, because hydrostatic equilibrium is well established in the time-averaged profiles, contributions from kinetic components (i.e., ram pressure) are nearly negligible.  This behavior is analogous to neutron star accretion columns \citep[e.g.,][]{Zhang2023}, which are likewise highly oscillatory in time but achieve a well-defined force balance when averaged over time.  A more detailed analysis of the current-sheet dynamics will be presented in subsequent papers, based on sub-Eddington models performed on our high-resolution grid. 

\subsubsection{Angular Momentum Transport}
\label{sec:angular_momentum_transport}

\begin{figure*}
    \centering
    \includegraphics[width=\textwidth]{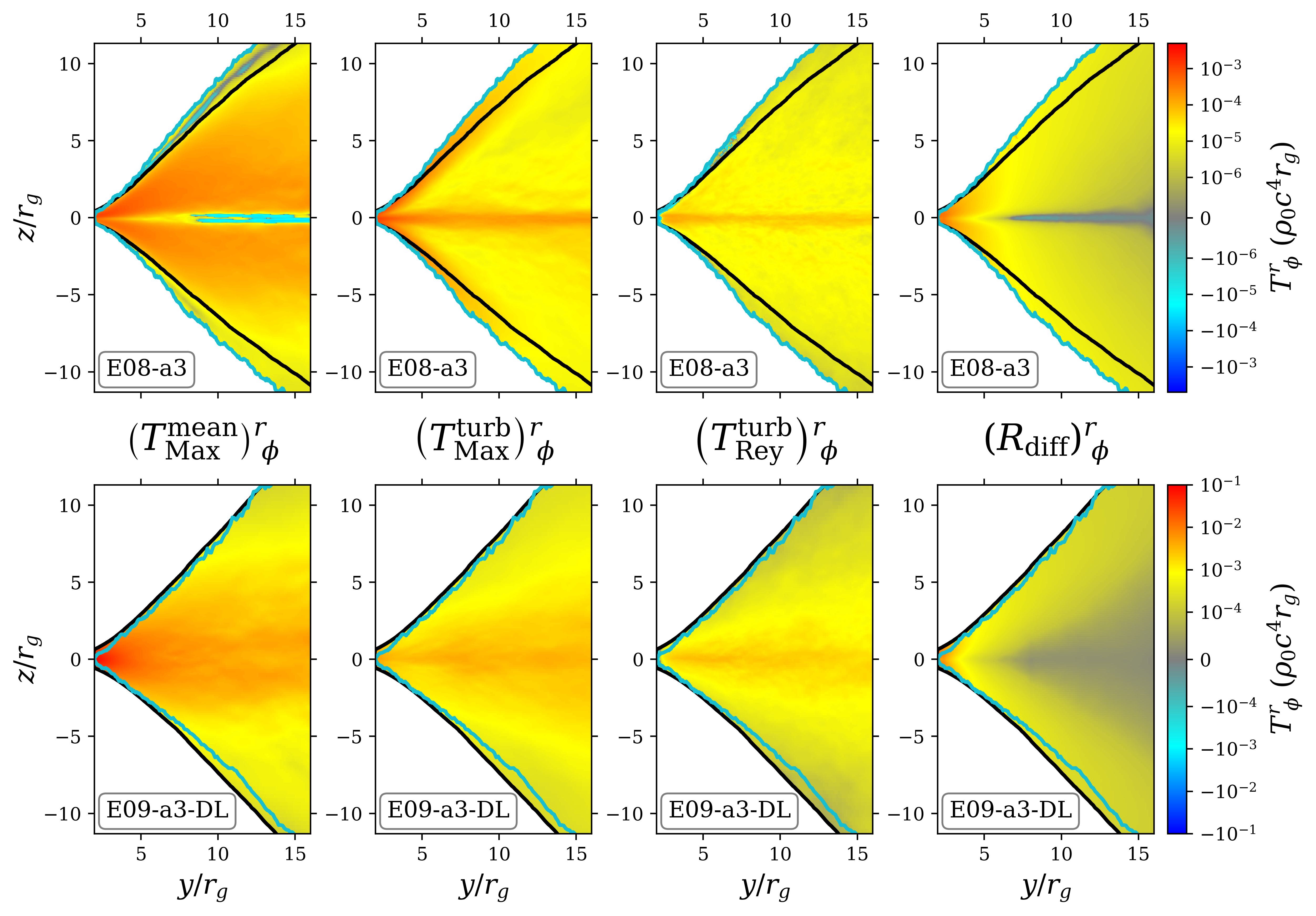}
    \caption{
    2D radial angular momentum flux components of two representative near-Eddington models in time- and azimuthal average.  The top row corresponds to the thin thermal disk embedded within a magnetic envelope, and the bottom row to the magnetically elevated disk.  From left to right column, the columns show the mean-field Maxwell stress, turbulent Maxwell stress, turbulent Reynolds stress, and radiation diffusion stress.  Auxiliary lines mark the scattering photosphere (black) and disk surface (cyan).  Detailed definitions of each component are provided in Section~3.5.1 of \citetalias{PaperII}. 
    } 
    \label{fig:angmom2d}
\end{figure*}

The accretion process in both near-Eddington steady-state solutions is strongly influenced by the mean magnetic field, as shown in \autoref{fig:angmom2d}, unlike the super-Eddington models, which are driven mainly by turbulence (see Section~3.5 in \citetalias{PaperII} for details).  Radiation plays a negligible role in both models, though its contribution increases at small radii near the black hole.  

In the thin thermal disk, although turbulent stresses dominate near the midplane and close to the disk surface, most of the accretion occurs within the magnetic envelope, where outward angular momentum transport is dominated by the mean-field Maxwell stress.  In the magnetically elevated disk, angular momentum transport is more homogeneous throughout the disk but increases smoothly toward the midplane and smaller radii, with mean-field Maxwell stress and turbulent stresses contributing comparably.  

The angular momentum transport in the double-loop model E07-a3-DL, which evolves from a magnetically elevated disk into a thin thermal disk, exhibits a transitional character. Although the disk eventually collapses into a thin thermal state, the turbulent component of angular momentum transport inherited from the elevated phase remains relatively strong.  This likely occurs because the accumulated vertical magnetic flux is sufficient to trigger collapse but insufficient to amplify the mean field through shear to the levels typically reached in single-loop thin thermal disk models.  As the evolution proceeds and the poloidal field continues to accumulate, differential rotation shears it into a stronger toroidal field, allowing the mean field to increase locally.  However, this progression is slow, and the simulation cannot be extended long enough to capture the eventual magnetic transition due to computational limitations. A more detailed discussion of this disk evolution is presented in \autoref{sec:magnetic_polarity_breaking}. 

\subsection{Outflows and Jet}
\label{sec:outflows_and_jet}

\begin{figure}
    \centering
    \includegraphics[width=\columnwidth]{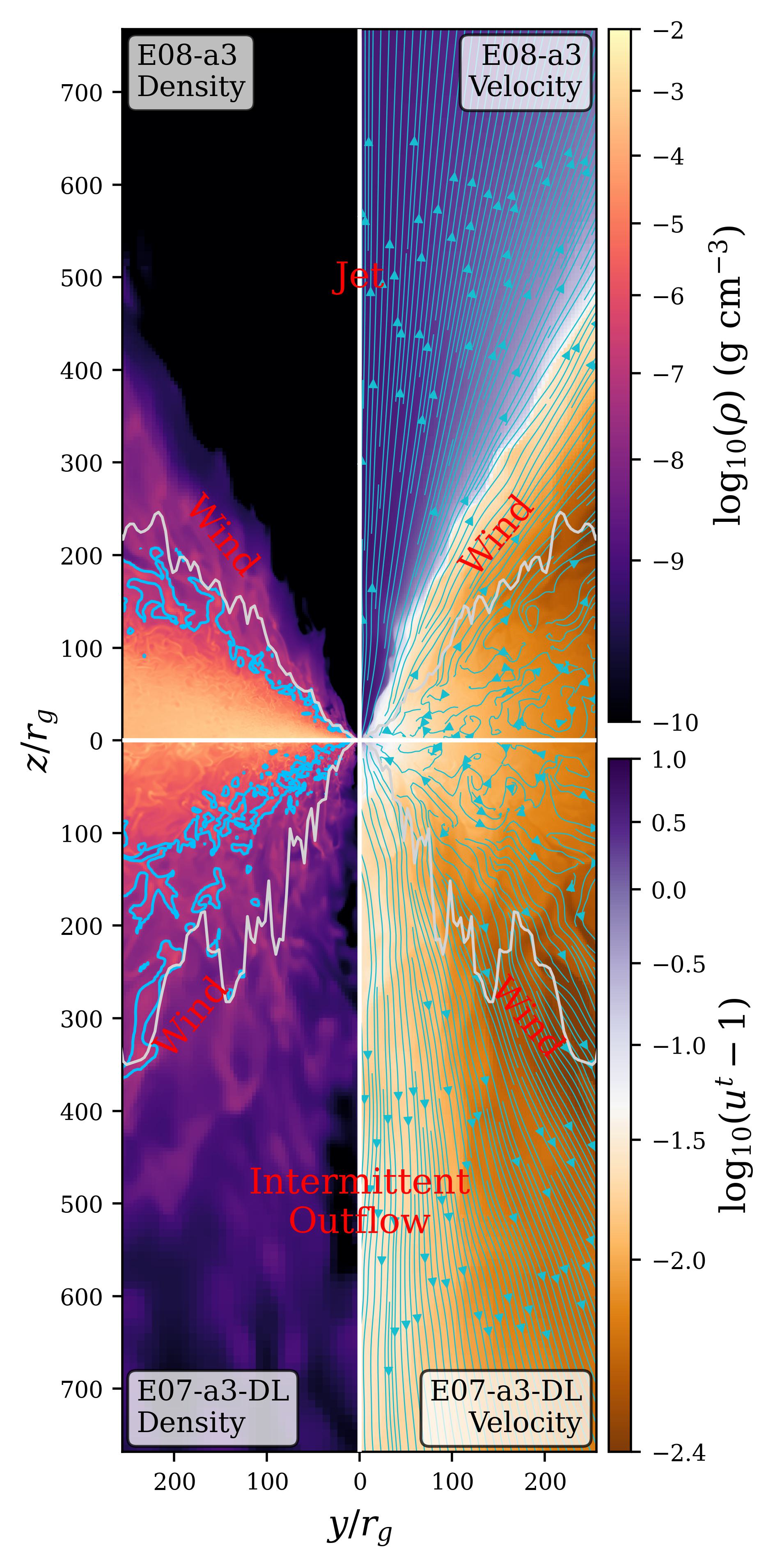}
    \caption{
    2D poloidal slices ($x=0$, $y>0$, $z>0$) illustrating outflows and jets in two representative near-Eddigton models, E08-a3 (top) and E07-a3-DL (bottom), which do and do not form jets, respectively.  Snapshots are taken after the systems have reached steady state: at $t=60000r_g/c$ for model E08-a3 and $t=70000r_g/c$ for model E07-a3-DL.  The left panels show gas density, while the right panels show the time component of the four-velocity (indicating fluid speed), with streamlines representing the velocity field.  In the right panels, the colormap highlights highly relativistic (purple), mildly relativistic (white), and non-relativistic (brown) flows.  Auxiliary lines mark zero Bernoulli parameter (blue contours) and the scattering photosphere (gray solid lines). 
    }
    \label{fig:outflow2d}
\end{figure}

\begin{figure}
    \centering
    \includegraphics[width=0.95\columnwidth]{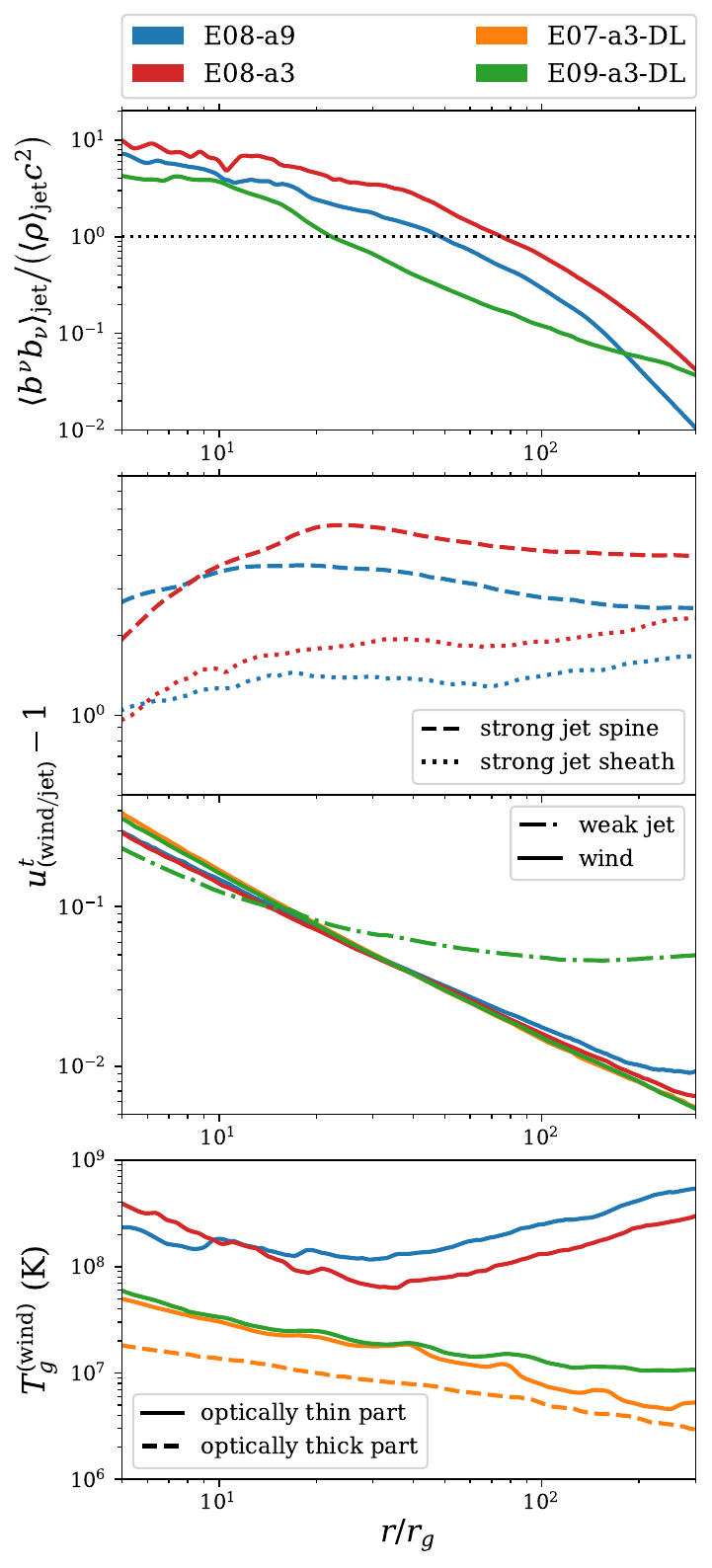}
    \caption{
    1D wind and jet profiles for all near-Eddington models.  From top to bottom, the panels show the jet magnetization parameter, jet and wind speeds, and wind temperature.  Model E07-a3-DL does not develop a persistent jet structure and therefore has no corresponding jet measurements.  \textbf{Top}: magnetization averaged within the jet region.  \textbf{Upper middle}: time component of the four-velocity averaged in the jet spine (dashed) and jet sheath (dotted) for models that form strong jets.  \textbf{Lower middle}: time component of the four-velocity averaged in the wind region (solid) and, where applicable, within a weak jet (dot-dashed).  \textbf{Bottom}: gas temperature averaged in the wind region, with the optically thick portion of the wind shown as a dashed line when present. All definitions of these measurements are provided in Section~3.6 of \citetalias{PaperII}. 
    }
    \label{fig:outflow1d}
\end{figure}

The jet region is defined by tracing velocity streamlines from the strongly magnetized funnel (where the magnetic energy exceeds the rest mass energy) that propagate to the edge of the computational domain (e.g., see Appendix~B in \citetalias{PaperII}).  The wind zone is then identified as the gravitationally unbound flow outside the jet region.  Using these definitions, all near-Eddington models produce powerful winds driven by a combination of radiation and magnetocentrifugal forces.  \autoref{fig:outflow2d} shows representative snapshots of the outflows and jets that arise in two selected models.  

\begin{figure*}
    \centering
    \includegraphics[width=\textwidth]{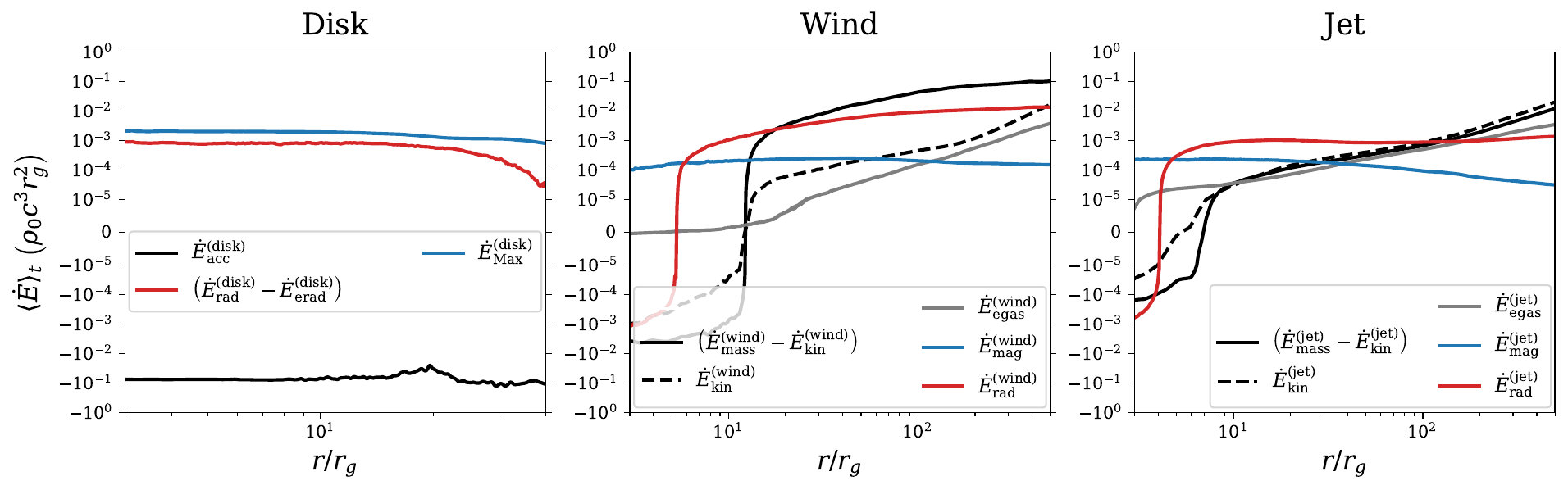}
    \caption{
    Energetic analysis of model E08-a9, based on time- and azimuthal averaged data.  From left to right, the panels show the energy partitioning of heating and cooling processes in the disk, wind, and jet regions.  All relevant definitions are provided in Appendix~E of \citetalias{PaperII}.  The remaining near-Eddington models (not shown for brevity) exhibit similar trends, with two exceptions: (1) in models E09-a3-DL, radiative cooling in the jet region dominates over all other forms of energy flux at large radii, and (2) in model E07-a3-DL, a well-defined jet region is absent.  Aside from these differences, the overall energetics are qualitatively consistent across these models. 
    }
    \label{fig:energy1d}
\end{figure*}

The winds are mildly relativistic and are launched near the disk surface in all near-Eddington models.  Among these cases, the two single-loop models, E08-a3 and E08-a9, launch strong and persistent jets, while the double-loop model E09-a3-DL develops an intermittent but coherent jet structure in time average.  In contrast, the double-loop model E07-a3-DL does not form a jet: the velocity streamlines originating in the strongly magnetized region near the black hole fail to reach the domain boundary and instead remains confined within $\sim 30r_g$ in the polar funnel.  As a result, only intermittent, weakly magnetized, mildly relativistic outflows appear in the funnel region, as shown in the bottom panels of \autoref{fig:outflow2d}. 

Similar to the behavior observed in the super-Eddington regime (see Section~3.6 in \citetalias{PaperII} for details), the strong jets are magnetically driven, whereas the weak, intermittent jet is powered by magnetic forces near the black hole but becomes radiation-driven at larger radii.  The 1D jet and wind profiles for all near-Eddington models are shown in \autoref{fig:outflow1d}.

As shown in the upper two panels, the strong jets undergo magnetic acceleration close to the black hole, with the spine reaching the peak velocity at $\sim 20r_g$ before gradually losing magnetic confinement.  Once the magnetization drops below unity (i.e., when the magnetic energy falls below the rest-mass energy) and the flow begins to freely stream, the jet spine expands laterally and merge into the jet sheath, as reflected by the corresponding increase in sheath velocity near $\sim 60r_g$. 

Unlike the strong jet, which is highly relativistic, the weak jet is intermittent and remains only mildly relativistic in time average, closely resembling the winds, as shown in the lower middle panel.  The mildly relativistic winds themselves generally follows a power-law velocity profile.  

The bottom panel shows the gas temperature averaged in the wind region.  Winds in most near-Eddington models are optically thin, except in model E07-a3-DL where a portion near the disk surface remains optically thick.  This optically thick component generally has a lower temperature than the optically thin region because stronger radiation-gas thermal coupling enables more efficient radiative cooling. 

Near the black hole, the wind is confined to a relatively narrow layer between the jet and the disk, so even though it is optically thin, it remains dense enough to stay moderately thermally coupled to the radiation, producing a power-law decline in temperature with radius.  Such hot, moderately optically thin winds may be related to the ultrafast outflows observed in luminous accreting black holes (e.g., \citealt{Tombesi2010,Walton2016}), which are generally inferred to originate from the inner disk regions.  In the two single-loop models, however, the wind temperature departs from this power law and begins to rise beyond $\sim 30r_g$, where the density becomes too low to maintain efficient gas-radiation coupling as the wind expands. 

\begin{deluxetable*}{l c c c c c c c c c}
\tablecaption{Comparison of cooling efficiencies and contributions from different cooling mechanisms \label{tab:energy_eql}}
\tablehead{
\colhead{Name} & 
\colhead{$\eta_{50}^{(\mathrm{rad})}$} & 
\colhead{$\eta_{50}^{(\mathrm{wind})}$} & 
\colhead{$\eta_{50}^{(\mathrm{jet})}$} & 
\colhead{$\eta_{200}^{(\mathrm{wind})}$} &
\colhead{$\epsilon_{\mathrm{kin, 50}}^{(\mathrm{wind})}$} & 
\colhead{$\epsilon_{\mathrm{mag, 50}}^{(\mathrm{wind})}$} & 
\colhead{$\epsilon_{\mathrm{kin, 50}}^{(\mathrm{jet})}$} & 
\colhead{$\epsilon_{\mathrm{mag, 50}}^{(\mathrm{jet})}$} & 
\colhead{$\epsilon_{\mathrm{floor, 50}}^{(\mathrm{jet})}$} 
\\
& \colhead{(\%)} & \colhead{(\%)}& \colhead{(\%)}& \colhead{(\%)}& \colhead{(\%)}
& \colhead{(\%)} & \colhead{(\%)}& \colhead{(\%)}& \colhead{(\%)}
\\
\quad\;\;\;(1) & (2) & (3) & (4) & (5) & (6) & (7) & (8) & (9) & (10)
}
\startdata
    E09-a3-DL  & 3.51 & 0.18 & 0.005 & 0.18 & 57 & 43 & 22 & 77 & 0.3 \\ 
    E08-a9     & 9.95 & 0.61 & 0.95  & 2.37 & 42 & 50 & 52 & 18 & 80  \\
    E08-a3     & 8.69 & 0.44 & 1.50  & 1.60 & 33 & 59 & 61 & 8  & 87  \\
    E07-a3-DL  & 5.26 & 0.47 & -     & 0.85 & 64 & 36 & -  & -  & -   \\
    \hline
\enddata
\tablecomments{
    The subscript indicates the radius at which each quantity is measured.  Relevant definitions are provided in Section~3.7 of \citetalias{PaperII}.  Measurements at $50r_g$ are used because the luminosities for all models have roughly stabilized at this radius. 
    {\bf Columns (from left to right):}  
    (1) Model name; 
    (2) Radiation efficiency; 
    (3) Wind efficiency; 
    (4) Jet efficiency;
    (5) Wind efficiency measured at a larger radius (where the wind power stabilizes);
    (6) Kinetic contribution to wind cooling;
    (7) Magnetic contribution to wind cooling;
    (8) Kinetic contribution to jet cooling;
    (9) Magnetic contribution to jet cooling;
    (10) Fraction of jet kinetic energy attributed to the density floor. 
}
\end{deluxetable*}

\subsection{Energy Flow and Radiation Production}

\subsubsection{Global Energetics}

All near-Eddington models, regardless of disk type, exhibit broadly similar energetics across the disk, wind, and jet regions, with only subtle differences within the disk between models employing single-loop versus double-loop magnetic field topologies.  Here, we use model E08-a9 as an example to illustrate the energy flow across the different partitions of the system.  The analysis procedure is established in Section~3.7 of \citetalias{PaperII}, and the definitions of the energy partitions in each region are provided in Appendix~E of \citetalias{PaperII}. 

As shown in \autoref{fig:energy1d}, the energy input comes from the disk accretion process, with a small fraction ($\lesssim 5\%$ of the accretion power) removed through the outgoing Poynting flux and radiation diffusion.  Most of the cooling occurs in the wind and jet regions through radiation emerging from the disk.  It is important to note, however, that this emergent radiation is only weakly coupled to the gas in these regions; consequently, the radiation flux there primarily reflects escaping energy from the disk rather than direct cooling of the wind or jet material. 

In the wind region, aside from the emergent radiation and the rest-mass energy carried by the outflow, cooling is dominated by the Poynting flux in the inner region ($\lesssim 50r_g$).  Further out, the kinetic energy flux gradually becomes dominant and eventually approaches the radiative flux at larger radii ($\gtrsim 500r_g$).  In the jet region, the overall behavior is similar to that of the wind, but the kinetic energy flux is substantially stronger.  

The only major differences in the remaining models arise from the absence of a strong jet in the double-loop cases.  In model E09-a3-DL, a weak jet forms and exhibits slightly enhanced radiative coupling due to higher gas density; it remains mildly relativistic and partially radiation-dominated, with Poynting flux dominating at small radii and radiation at large radii, while the kinetic component never becomes significant.  Model E07-a3-DL does not even develop a well-defined jet region and is therefore excluded from the jet analysis. 

\subsubsection{Comparison across Simulations}

In \autoref{tab:energy_eql}, we summarize the cooling efficiency by system partition (wind and jet) and by physical contribution (radiative, magnetic, and kinetic), with the relevant definitions provided in Section~3.7 of \citetalias{PaperII}.  Regardless of magnetic topology, radiative cooling dominates in all near-Eddington models, in contrast to the super-Eddington cases where outflow cooling can be comparable to, or even exceed, radiative cooling.  

Compared with the double-loop models, the single-loop models produce much more powerful winds, most likely due to a stronger Blandford-Payne \citep{Blandford1982} process.  Although it is difficult to quantify the relative roles of radiation and magnetic fields in driving the outflows, the single-loop models clearly shows a larger magnetic contribution.  

The wind power correlates positively with the vertical magnetic flux: higher black hole spin (E08-a9 vs. E08-a3) and greater vertical flux (single-loop vs. double-loop) both lead to stronger winds.  Interestingly, because model E07-a3-DL accumulates vertical magnetic flux in the inner disk during its evolution (polarity breaking; see \autoref{sec:magnetic_polarity_breaking} for discussion), its wind power lies between that of the single-loop cases and the double-loop model E09-a3-DL, which remains magnetically elevated without polarity breaking. 

\begin{figure*}
    \centering
    \includegraphics[width=\textwidth]{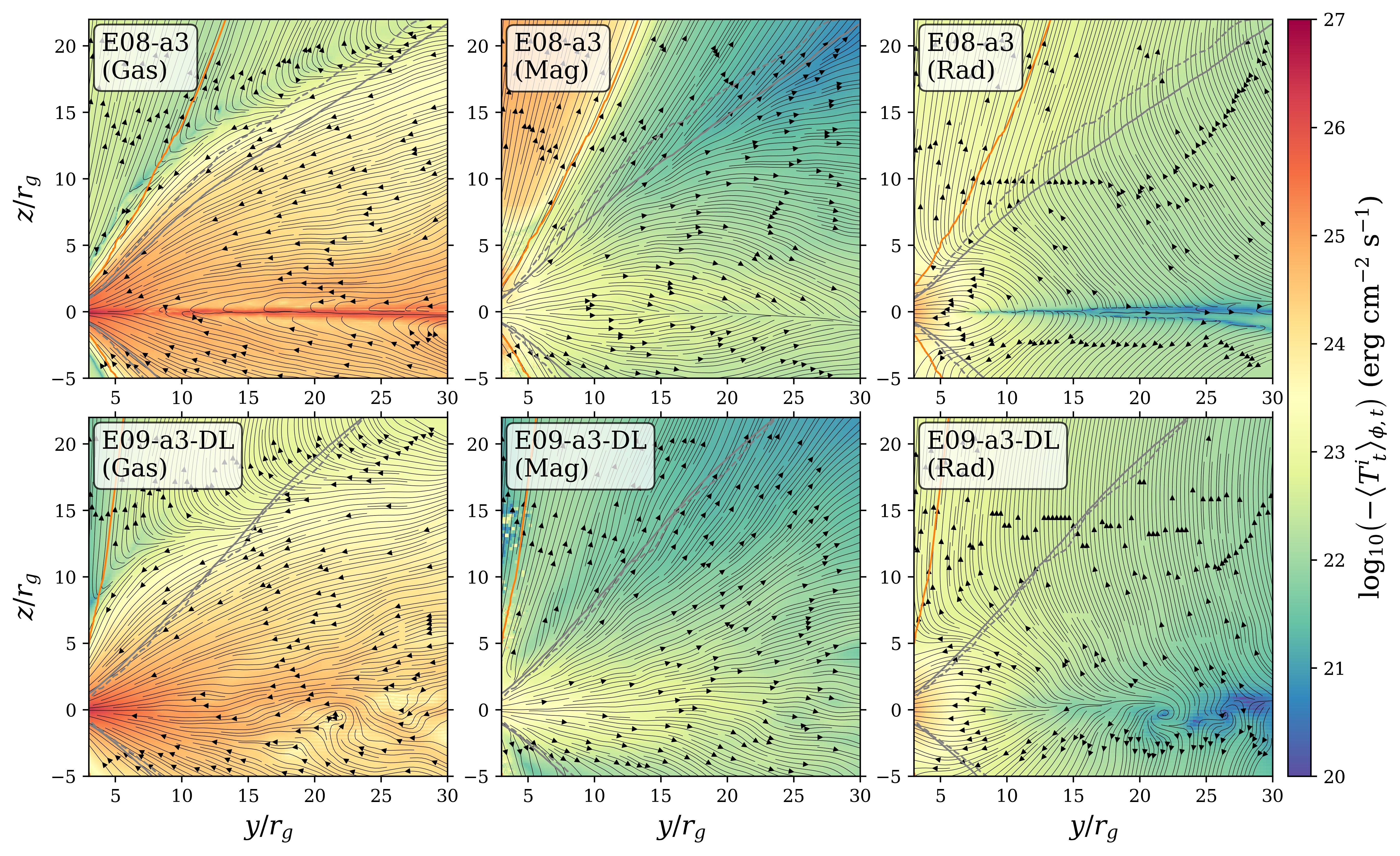}
    \caption{
    2D profiles of the poloidal energy flux of two representative near-Eddington models, shown in time- and azimuthal average.  The top row corresponds to the thin thermal disk embedded within a magnetic envelope, and the bottom row to the magnetically elevated disk.  From left to right, the columns show the magnitude (colormap) and direction (streamlines) of the energy flux carried by gas, magnetic fields, and radiation.  Auxiliary lines indicate the scattering photosphere (solid gray), the disk surface (dashed gray), and the jet boundary (orange).  Detailed definitions of each component are provided in equation~\autoref{eq:energy_flux_partition}. 
    }
    \label{fig:energy_flow2d}
\end{figure*}

The trend in jet power is consistent with the analysis in \autoref{sec:outflows_and_jet}.  However, the measurements of the strong jet can only be interpreted qualitatively because they are dominated by the density floor.  In contrast, the weak jet in model E09-a3-DL is barely affected by the density floor.  The jet is baryon-loaded, and its the kinetic contribution is much smaller than the magnetic one.  This reflects its mildly relativistic speed rather than strong magnetization.  Indeed, as shown in \autoref{fig:profile2d}, the magnetization of the weak jet is significantly lower than that of the strong jet.

\subsubsection{Disk Energy Transport}

\autoref{fig:energy_flow2d} presents a side-by-side comparison of the energy flow in the two near-Eddington disk types.  The total energy flux is decomposed into gas, magnetic, and radiation contributions following equation~\autoref{eq:energy_flux_partition}.  Within the disk, the gas component is dominated by the rest-mass energy associated with the accretion flow, the magnetic component by the Maxwell stress, and the radiation component primarily by diffusion. 

\begin{subequations}    
\begin{align}
    &\left(T^i_{\ t}\right)_{\mathrm{gas}} = \left(\rho + \frac{\gamma}{\gamma-1}P_g\right) u^i u_t
    \ ,
    \\
    &\left(T^i_{\ t}\right)_{\mathrm{mag}} = b^{\nu}b_{\nu}u^i u_t - b^i b_t
    \ ,
    \\
    &\left(T^i_{\ t}\right)_{\mathrm{rad}} = R^i_{\ t}
    \ .
\end{align}
\label{eq:energy_flux_partition}
\end{subequations}

Although the overall cooling behavior is broadly similar across regions in all near-Eddington models, the energy flow within the accretion disk differs significantly between the thin thermal disk and the magnetically elevated disk.  In the magnetically elevated disk, the energy flow pattern is relatively straightforward: gas inflow occurs throughout the disk, driving accretion and heating the system via turbulence; magnetic energy is carried outward in a roughly radial direction by the Poynting flux; and radiation escapes nearly perpendicular to the disk, primarily through diffusion.  Only in the very inner region ($\lesssim 10r_g$) can radiation become trapped in the inflow and be advected inward. 

In the thin thermal disk, the accretion power delivered to the black hole is extracted by the Poynting flux as magnetic energy and eventually released as radiation emerging from the disk.  Accretion proceeds mainly through the magnetically dominated envelope above the midplane, whereas the midplane region accretes much less efficiently and can even transition to decretion beyond $10r_g$.  Magnetic energy transported outward from the inner disk partially converges toward the midplane and dissipates as heat, where a global current sheet develops, making this region highly dissipative. The resulting radiation, generated primarily in this dense midplane layer, escapes through the magnetic envelope and emerges roughly perpendicular to the disk surface.  Within the optically thick envelope, the escaping flux is slightly tilted inward due to radiation advection.  

\subsubsection{Radiation Production}

\begin{figure}
    \centering
    \includegraphics[width=\columnwidth]{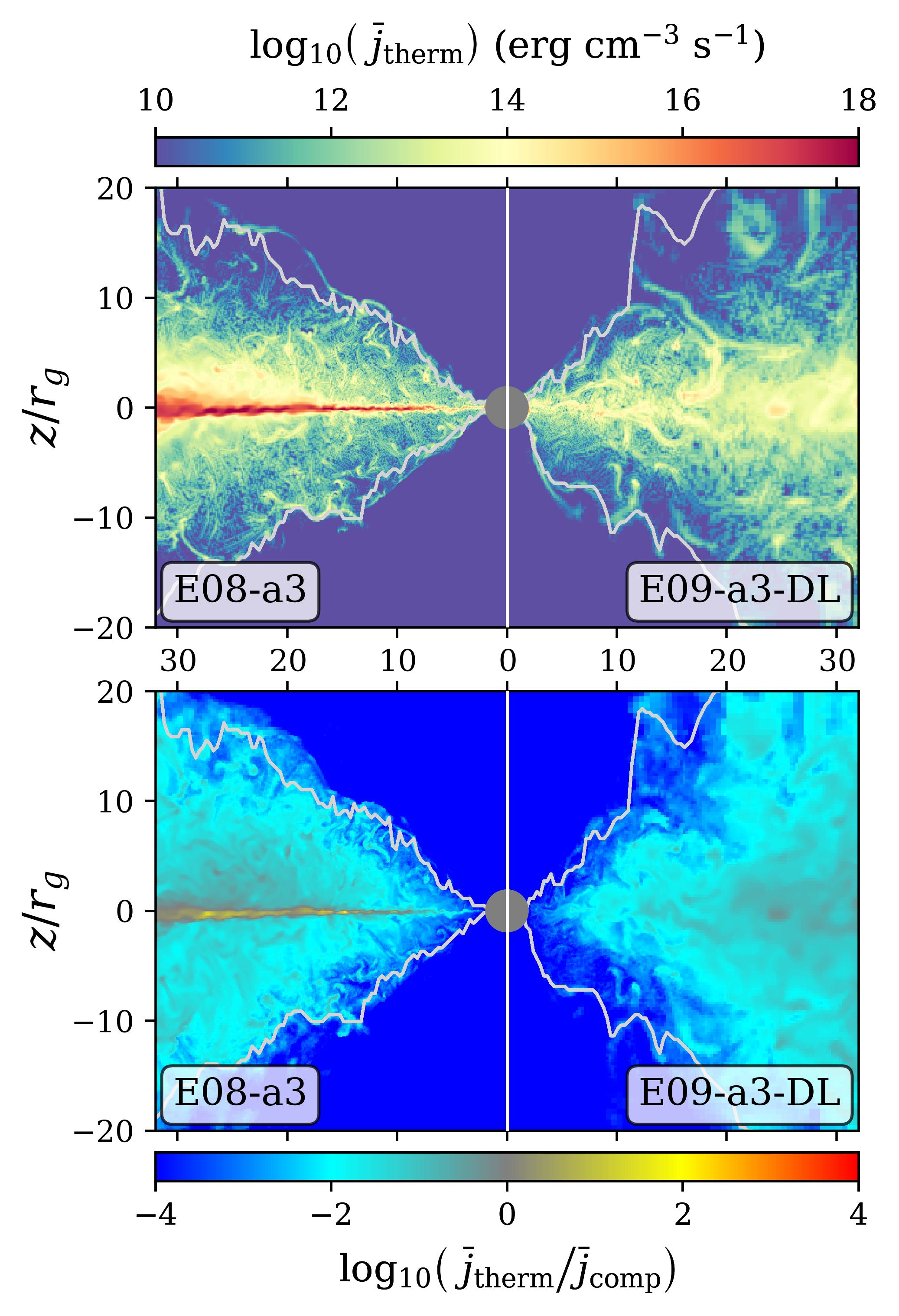}
    \caption{
    Thermal emissivity comparison for two representative near-Eddington models.  Snapshots at $t=60000r_g/c$ show the thin thermal disk within a magnetic envelope (left), and the magnetically elevated disk (right).  The upper panel displays the thermal emission term, and the lower panel shows the ratio of thermal emission to thermal Comptonization.  Definitions of these quantities are provided in equation~\autoref{eq:rad_em}.  Gray lines indicate the scattering photosphere.  
    }
    \label{fig:rad_src2d}
\end{figure}

Radiation is produced through thermal emission and can be further energized by thermal Comptonization.  The radiation source term follows equation~(4) of \citetalias{PaperI}, where thermal emission and Comptonization are the only processes that exchange internal energy between gas and radiation, defined as follows: 
\begin{subequations}    
\begin{align}
    &\bar{j}_{\mathrm{therm}} = \rho\kappa_P\left(\frac{a_r T_g^4}{4\pi} - \bar{J}\right)
    \ ,
    \\
    &\bar{j}_{\mathrm{comp}} = \rho\kappa_T\frac{4k_B(T_g-T_r)}{m_ec^2}\bar{J}
    \ . 
\end{align}
\label{eq:rad_em}
\end{subequations}
\autoref{fig:rad_src2d} presents the thermal emission (where radiation is generated) and its ratio to Comptonization (how radiation is energized) in a side-by-side comparison of the two disk types.  In the thin thermal disk, photons are predominantly produced in the dense midplane layer through magnetic heating, while thermal Comptonization just above the midplane heats them to a level comparable to the thermal emission and becomes the dominant energy-exchange mechanism near the disk surface.  In contrast, the magnetically elevated disk generates radiation more uniformly throughout its main body, with comparable Compton heating at moderate heights; however, towards the disk surface or near the black hole, Comptonization becomes significantly stronger, indicating much more efficient photon hardening than in the thin thermal disk.

\section{Discussion} 
\label{sec:discussion}

\subsection{Magnetic Polarity Breaking}
\label{sec:magnetic_polarity_breaking}

\begin{figure*}
    \centering
    \includegraphics[width=\textwidth]{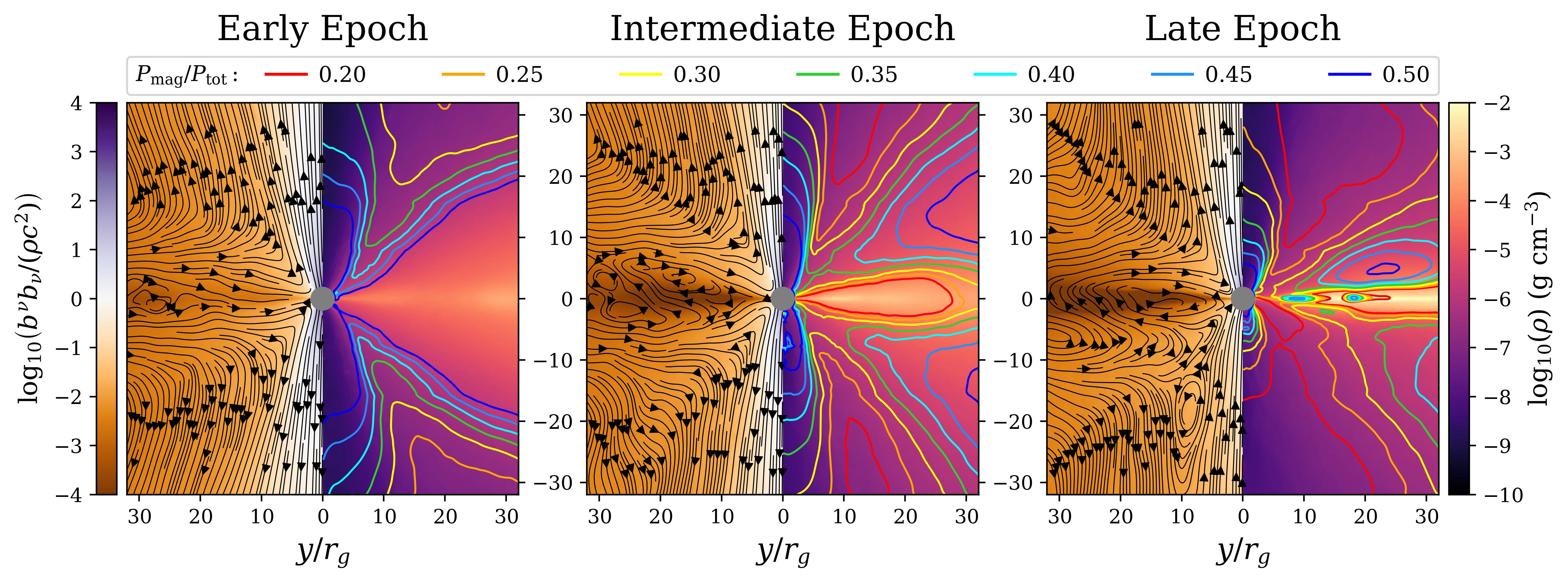}
    \caption{
    Disk evolution in model E07-a3-DL from a magnetically elevated disk to a thin thermal disk.  The 2D profiles are shown in time- and azimuthal average over three epochs within a time window of $\Delta t=3000r_g/c$: early ($10000-13000 r_g/c$), intermediate ($17000-20000 r_g/c$), and late ($57500-60500 r_g/c$), from left to right.  Each panel shows magnetization with magnetic field streamlines on the left, and gas density with contours of the magnetic-to-total pressure ratio on the right.  Pressure ratio contours progress from magnetically dominated (blue) to thermal dominated (red), as indicated by the legend. 
    }
    \label{fig:trans}
\end{figure*}

\begin{figure}
    \centering
    \includegraphics[width=\columnwidth]{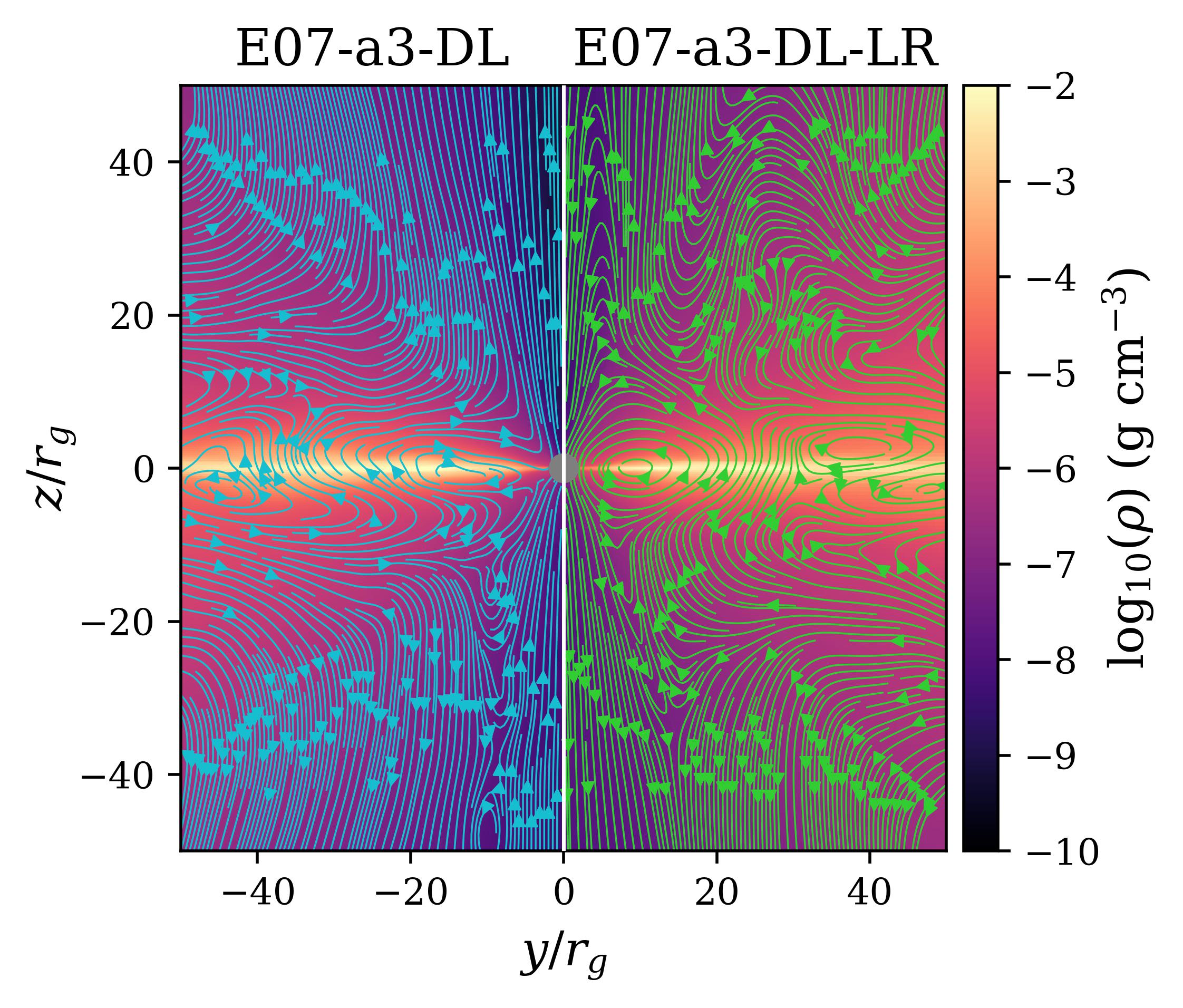}
    \caption{
    Side-by-side comparison of the final states of model E07-a3-DL at intermediate (left) and low (right) resolution.  Density and magnetic field quantities are time- and azimually averaged after the system reaches a steady state: $t=57500$-$67500r_g/c$ for the intermediate-resolution model and $t=50000$-$60000r_g/c$ for the low-resolution model.  Magnetic field streamlines are plotted in different colors for clarity.  The magnetic polarity in the final state happens to be opposite in the two cases, reflecting the inherently stochastic nature of the transition driven by radiation-powered outflows. 
    }
    \label{fig:trans_res}
\end{figure}

An interesting feature of the double-loop model E07-a3-DL is its evolution from a magnetically elevated disk to a thin thermal disk, driven by the gradual buildup of vertical magnetic flux near the midplane (see the right column of \autoref{fig:timestream}).  \autoref{fig:trans} presents the time- and azimuthally averaged 2D profiles tracing this transition.  

In the early epoch, shortly after the system relaxes from the initial condition, the flow becomes magnetically elevated and accretes at a slightly super-Eddington rate ($\sim 1$-$2$; see the green curves in \autoref{fig:hst_compare}).  During this phase, magnetic pressure becomes increasingly dominant toward the midplane, forming a magnetically supported structure similar to model E09-a3-DL (though emerging from a somewhat lower accretion rate).  This state is short-lived: the elevated accretion rate drives a strong, radiation-powered outflow that stochastically and anisotropically removes magnetic fields in the inner disk.  As a result, the magnetic flux ratio between the southern and northern disk surfaces gradually increases from nearly zero to roughly three.  

During the intermediate epoch, as magnetic fields are stripped unevenly from the two disk surfaces, compensating fields are induced near the midplane.  Vertical magnetic flux gradually accumulates in this region, causing the magnetically supported structure to collapse into a thin disk.  A current sheet subsequently develops near the midplane, accelerating the lose of magnetic support, while the peak of the magnetic energy density shifts above the midplane, further compressing the disk.  

In the late epoch, the flow settles into a geometrically thin, thermally supported disk, overlaid by a magnetized envelope.  Midplane symmetry in the magnetic field breaks, producing two counter-clockwise magnetic loops; however, the loop closer to the black hole is preferentially accreted.  Consequently, the magnetic flux threading the black hole becomes single-polarity (upward in this case).  The resulting thin thermal disk closely resembles those in the single-loop configurations, though its overlying envelope is less magnetically dominated, likely because the vertical magnetic flux that accumulates near the midplane is comparatively weaker.  

This magnetic polarity breaking mechanism is inherently stochastic.  As shown in \autoref{fig:trans_res}, both the intermediate- and low-resolution models undergo the same disk transition.  Although they converge to similar disk geometries and magnetic topologies, they end up with opposite magnetic polarities near the midplane, reflecting the stochastic nature of the polarity-breaking process. 

Recall that the single-loop and double-loop configurations were designed to probe how the system behaves under maximized and minimized vertical magnetic flux.  However, the transition observed in model E07-a3-DL suggests that, in a high-accretion regime, the thin thermal disk is a more likely outcome than the magnetically elevated configuration.  This preference can be attributed to two factors: 
\begin{enumerate}
    \item Even a modest amount of vertical magnetic flux is sufficient to trigger the formation of a midplane current sheet and drive the flow toward a thin thermal disk. 
    \item In a rapidly accreting system, it is difficult to maintain the weak or near-zero vertical flux required to sustain a magnetically elevated disk.  Stochastic outflows driven by radiation feedback can disrupt magnetic polarity and facilitate the accumulation of net vertical magnetic flux. 
\end{enumerate}

This interpretation also explains why model E09-a3-DL, despite starting from the same double-loop magnetic topology, does not undergo a similar transition.  Its lower initial accretion rate produces weaker radiation feedback, which is insufficient to significantly reorganize the magnetic field topology. As a result, the initial weak-flux configuration is largely preserved, allowing the magnetically elevated disk to persist throughout the entire simulation.

\subsection{Trends with Spin and Disk Type}

In this section, we examine how black hole spin and disk type affect the accretion system in the radiation-dominated regime, and provide a qualitative assessment of their influence on disk structure and accretion dynamics.  We begin by comparing the spin dependence of outflow power and angular momentum transport in two near-Eddington models (E08-a3 and E08-a9) and two super-Eddington models (E88-a3 and E150-a9; see \citetalias{PaperII} for details).  
We then summarize trends with spin and disk type across the non-radiative, near-Eddington, and super-Eddington cases. 

\begin{figure}
    \centering
    \includegraphics[width=\columnwidth]{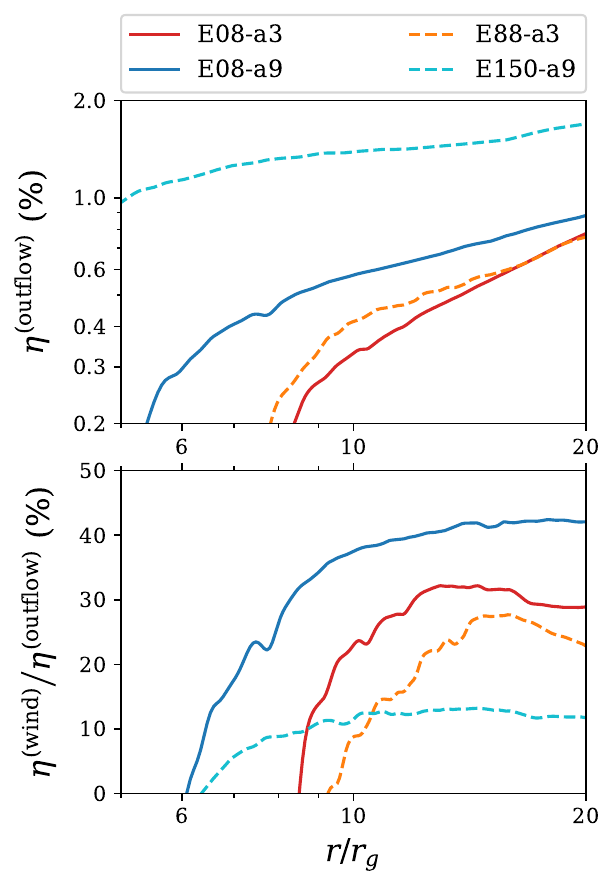}
    \caption{
    Spin dependence of outflows in the near- and super-Eddington accretion regimes.  The outflow includes both jet and wind components, with their cooling efficiencies defined in Section~3.7.3 of \citetalias{PaperII}.  For each accretion regime (solid lines for near-Eddington and dashed lines for super-Eddington), two representative models are shown: a low-spin case ($a=0.3$, red/orange) and a high-spin case ($a=0.9375$, blue/cyan).  The top panel presents the radial profiles of the total outflow efficiency, while the bottom panel shows the percentage contribution of the wind component. 
    }
    \label{fig:spin_dep_outflows}
\end{figure}

Black hole spin can affect the dynamics 
primarily through its interaction with magnetic fields.  In the inner disk, the spinning black hole drags spacetime and twists the magnetic field lines, powering outflows in the form of jets and winds through the Blandford-Znajek \citep{Blandford1977} and Blandford-Payne \citep{Blandford1982} processes, respectively.  In the main disk body, spin can influence 
the magnetic field structure and therefore the Maxwell stress, which dominates angular momentum transport and drives the accretion flow, 
potentially affecting the radial structure of the disk.  In addition to these magnetic effects, spin also alters the Kerr metric near the black hole and can modify the equilibrium conditions of the flow, particularly its vertical structure. 

\autoref{fig:spin_dep_outflows} shows how the total outflow power (jet + wind) depends on black hole spin in both the near- and super-Eddington accretion regimes.  
Among the models compared here, the higher spin cases produce stronger outflows at comparable accretion rates, though the underlying mechanisms differ between the two regimes.  In the super-Eddington case, higher spin primarily strengthens the jet: the radiation-supported thick disk provides effective collimation, and the jet contributes roughly $70$-$90$\% of the total outflow power.  In contrast, in the near-Eddington case, the spin dependence emerges mainly in the wind component.  For instance, from models E08-a3 to E08-a9, the wind contribution increases by $\sim10$\%.  This arises because a thin disk does not effectively collimate the jet, allowing the twisted magnetic fields to remain less horizontally confined and thereby redirect a larger fraction of the energy into the wind.

\begin{figure}
    \centering
    \includegraphics[width=\columnwidth]{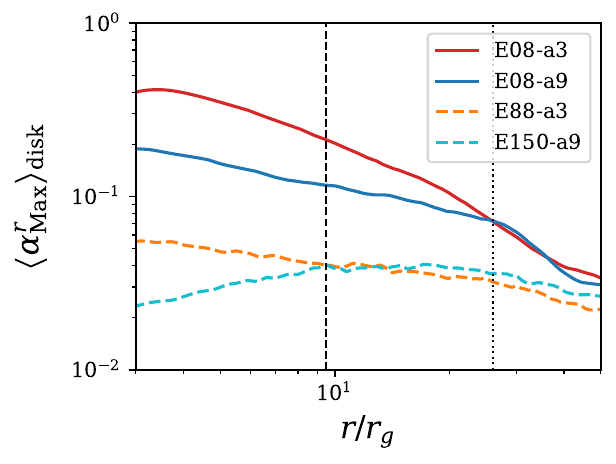}
    \caption{
    Spin dependence of the Maxwell component of the effective viscosity in the near- and super-Eddington accretion regimes.  The definition of this quantity is provided in Section~3.5.2 of \citetalias{PaperII}.  For each regime (solid lines for near-Eddington and dashed lines for super-Eddington), we show disk-averaged radial profiles for two representative models: a low-spin case ($a=0.3$, red/orange) and a high-spin case ($a=0.9375$, blue/cyan).  In both regimes, the high- and low-spin models exhibit distinct radial behavior in the inner disk but gradually converge to similar trends at larger radii.  Vertical dotted and dashed lines indicate the outermost radii where 
    the two spin cases show noticeable differences} in the near-Eddington and super-Eddington disks, respectively. 
    \label{fig:spin_dep_alpha}
\end{figure}

The accretion process can be characterized by the effective viscosity (or the pressure-scaled angular momentum flux; see Section~3.5.2 in \citetalias{PaperII} for details), which quantifies the efficiency of outward angular momentum transport.  \autoref{fig:spin_dep_alpha} shows how black hole spin influences the accretion process primarily through the Maxwell component of the effective viscosity.  Across both accretion regimes, the radial profiles exhibit strong differences between the two spins near the black hole but gradually converge to similar trends at larger radii, indicating that the outer disk is no longer sensitive to the spin.  

It is worth noting that this convergence occurs at smaller radii in the super-Eddington models than in the near-Eddington models.  This behavior is likely tied to intrinsic differences in their magnetic properties: super-Eddington disks are geometrically thick and dominated by turbulent magnetic fields, whereas near-Eddington disks are thinner and governed primarily by the mean-field component.  Because the mean magnetic field preserves its coherence over larger distances, it can carry spin-imprinted stresses farther out in the disk, while the strong turbulence in the super-Eddington disks rapidly disrupts this coherence, limiting the radial extent over which the spin influence persists. 

\begin{figure}
    \centering
    \includegraphics[width=\columnwidth]{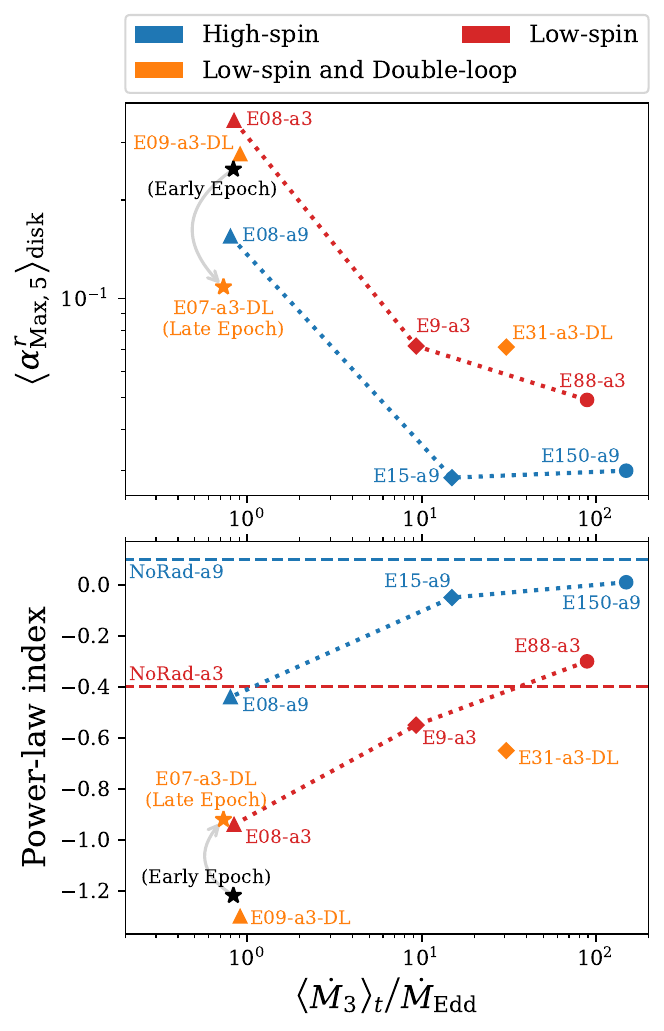}
    \caption{
    Accretion process as a function of accretion rate, black hole spin, and initial magnetic topology.  \textbf{Upper panel}: Radial profile of the Maxwell component of the effective viscosity.  \textbf{Lower panel}: Radial power-index of this component in the inner disk region.  All measurements are taken during the steady-state epoch and are averaged over time and azimuth within the disk region.  Non-radiative models are shown as horizontal lines independent of accretion rate because they are scale-free.  We also include an additional measurement from the early epoch ($t=10000$-$13000r_g/c$) of model E07-a3-DL, which evolves from a magnetically elevated disk to a thin thermal disk; this point is marked by the black star.  Blue and red markers denote high-spin and low-spin models initialized with single-loop magnetic fields, while orange markers represent low-spin models initialized with double-loop magnetic fields. 
    }
    \label{fig:spin_dep}
\end{figure}

To characterize the accretion process across the non-radiative, near-Eddington, and super-Eddington regimes, we analyze the radial Maxwell component of the effective viscosity, which dominates the angular momentum flux and follows a power-law profile in the inner disk (i.e., $\alpha \propto r^{\Gamma}$).  For demonstration, \autoref{fig:spin_dep} reports the disk-averaged effective viscosity at $r=5r_g$ along with the corresponding radial scaling from the fitted power-law index $\Gamma$.  

In our simulations, high black hole spin is associated with reduced efficiency of outward angular momentum transport through its magnetic influence, thereby weakening the accretion process.  Consequently, achieving the same accretion rate requires a higher gas density in the high-spin models.  This reduction in angular momentum transport 
may arise from MRI suppression in the more compact inner region at high spin, which weakens the generation of radial magnetic fields and their subsequent shearing into toroidal fields.  This spin effect also persists more uniformly at high spin, as reflected in the flatter radial dependence of the effective viscosity (with a power-law index approaching zero corresponding to a nearly constant profile).  

Across different mass accretion regimes, the disk develops distinct geometric structures: a thick disk in the non-radiative and super-Eddington models, and a thin disk in the near-Eddington models.  The non-radiative disks are thick because they are adiabatic, while the super-Eddington disks remain thick due to their low radiative efficiency (see Section~3.7.3 in \citetalias{PaperII}).  In contrast, the near-Eddington disks are geometrically thinner as a result of more efficient radiative cooling; depending on the available vertical magnetic flux, they form either a thin thermal disk or a magnetically elevated disk.  These structural differences strongly influence the underlying magnetic properties (turbulent or mean-field dominance) and therefore regulate angular momentum transport (e.g., relative dominance of turbulent versus mean-field stresses discussed earlier).  

\begin{figure*}
    \centering
    \includegraphics[width=\textwidth]{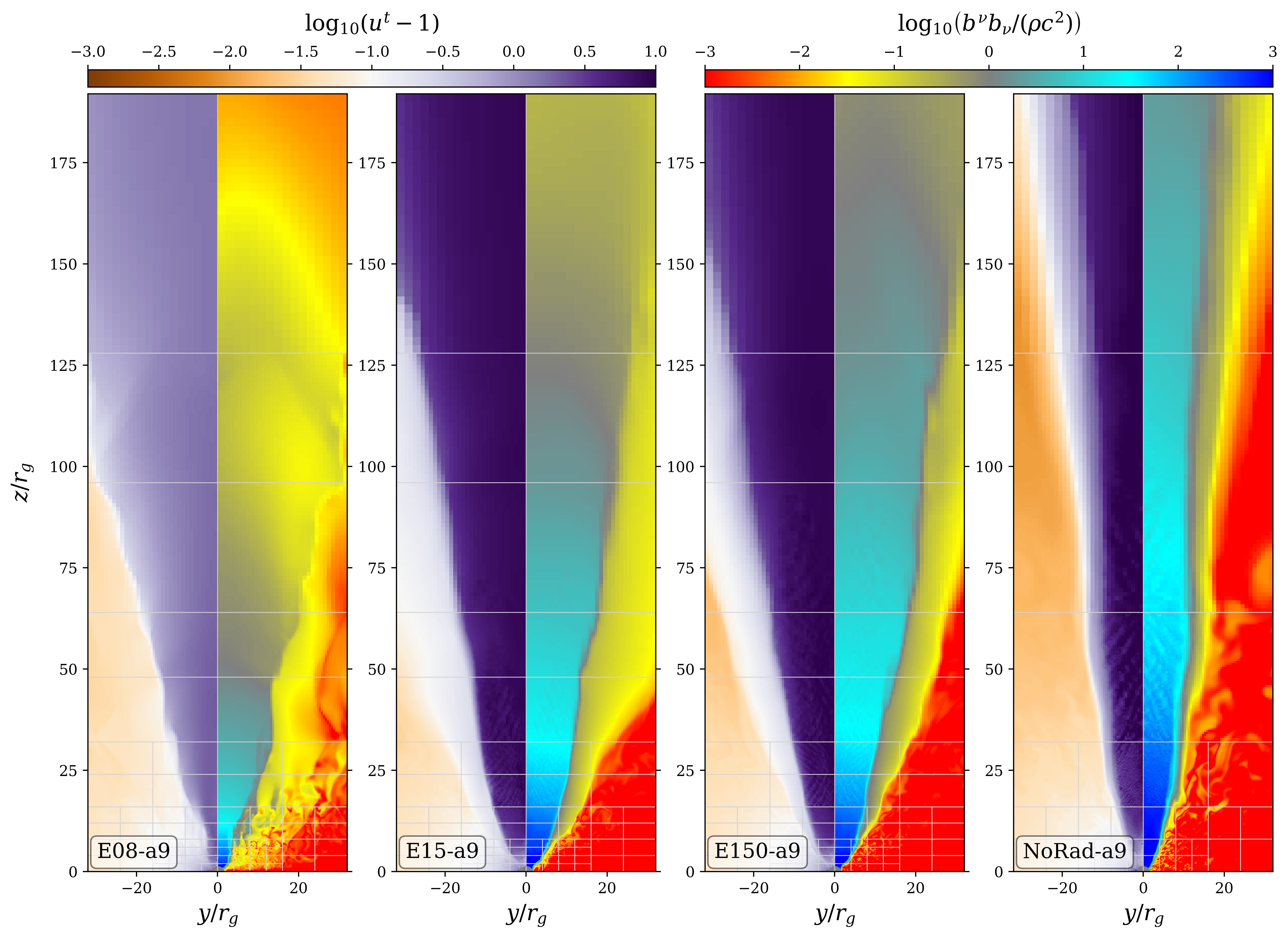}
    \caption{
    2D profiles of relativistic jets shown in order of increasing jet efficiency across representative high-spin models.  From left to right, snapshots are taken at $t=60000r_g/c$ for models E08-a9, E15-a9, and E150-a9 and at $t=34000r_g/c$ for model NoRad-a9.  In each panel, the left half displays the time component of the four-velocity, with a colormap distinguishing non-relativistic (orange), mildly relativistic (white), and highly relstivistic (purple) regions; the right half shows the jet magnetization, with a colormap highlighting strongly magnetized (blue/cyan), weakly magnetized (gray/yellow), and gas-dominated (red) regions.  Jet power increases with stronger disk collimation, with the collimating disk roughly identified by the gas-dominated regions (red) in the right-half panels. 
    }
    \label{fig:jet_compare}
\end{figure*}

Regardless of the accretion regime, thin disks generally remove angular momentum more efficiently than thick disks (i.e., higher effective viscosity), even though their efficiency declines more rapidly with radius (i.e., smaller power-law indices).  Comparing the thin thermal disk with the magnetically elevated disk, both show similar levels of Maxwell-stress-driven accretion, but the latter exhibits a steeper decline at larger radii, likely reflecting its slightly higher level of turbulent magnetic fields (see \autoref{sec:angular_momentum_transport}). 

In particular, the transition in model E07-a3-DL from a magnetically elevated disk to a thin thermal disk largely follows the trends seen in the other models.  During the early epoch, the disk is magnetically elevated and therefore resembles model E09-a3-DL, showing excellent agreement in both the magnitude of the Maxwell stress and its radial power-law scaling.  Once the disk collapses into a thin thermal structure in the late epoch, the power-law index shifts toward the value found in model E08-a3, which consistently represents a thin thermal disk.  However, the Maxwell viscosity itself becomes lower than in all other near-Eddington cases.  

This difference is likely related to the relatively low vertical magnetic flux in model E07-a3-DL.  Even after the collapse, the turbulent Maxwell stress remains relatively high (only slightly below the mean-field component), unlike in the single-loop model E08-a3, where the mean field clearly dominates (e.g., see \autoref{sec:angular_momentum_transport}).  In model E08-a3, the mean field is rapidly amplified through shear, allowing it to govern the disk dynamics.  In contrast, although the accumulated vertical magnetic flux in model E07-a3-DL is sufficient to trigger disk collapse, its relatively small magnitude requires more time for substantial mean-field amplification, delaying the transition to mean-field dominance.  Consequently, the relatively low Maxwell viscosity in model E07-a3-DL reflects this incomplete transition in magnetic dominance. 

\subsection{Jet Propagation and Disk Collimation}
Relativistic jets are launched through the Blandford-Znajek process \citep{Blandford1977}, which depends sensitively on the black hole spin and the strength of the poloidal magnetic flux threading the horizon.  However, the ability of these jets to sustain themselves and remain observable as they propagate outward depends critically on how effectively they are collimated, particularly near the jet base.  These jets generally begin to free-stream once their magnetization drops below unity (see \autoref{fig:outflow1d} or Figure~11 in \citetalias{PaperII}).  At that stage, they lose lateral confinement and expand sideways, and the distinction between the jet spine and sheath gradually disappears as the two merge. 

In our models that produce relativistic jets, we find that jet propagation is strongly influenced by disk collimation.  Effective collimation not only laterally confines the outflow, facilitating its propagation, but also enable more efficient acceleration to highly relativistic speeds by concentrating the magnetic forces within a small solid angle.  \autoref{fig:jet_compare} presents the jet profiles for the non-radiative, near-Eddington, and super-Eddington models, arranged in order of increasing jet efficiency.  

In the non-radiative model, the accretion flow is adiabatic and forms the geometrically thickest disk, providing the strongest collimation among all cases.  The resulting jet is tightly collimated, remains strongly magnetized out to large radii ($\sim300r_g$), and reaches highly relativistic velocities within the narrowest funnel opening angle. 

In the super-Eddington regime, the accretion disk is thermally expanded and remains geometrically thick.  Although radiative cooling is present, the small funnel limits the radiation cooling efficiency, keeping the inflow nearly adiabatic close to the black hole (e.g., see Section~3.2 in \citetalias{PaperII}).  As a result, the jet is again well collimated by the thick inner disk, similar to the non-radiative case.  However, as the funnel opens slightly at lower accretion rates, the jet becomes less collimated, and the strongly magnetized region correspondingly contracts in radius (e.g., from $\sim175r_g$ in model E150-a9 to $125r_g$ in model E15-a9). 

In the near-Eddington regime, the inner accretion disk becomes too geometrically thin to provide effective collimation.  A relativistic jet still forms, but its power is substantially reduced, and the jet develops a much wider opening angle due to the lack of lateral confinement.  The strongly magnetized portion of the jet also contracts to within only $\sim 50r_g$ of the black hole. 

\begin{figure}
    \centering
    \includegraphics[width=\columnwidth]{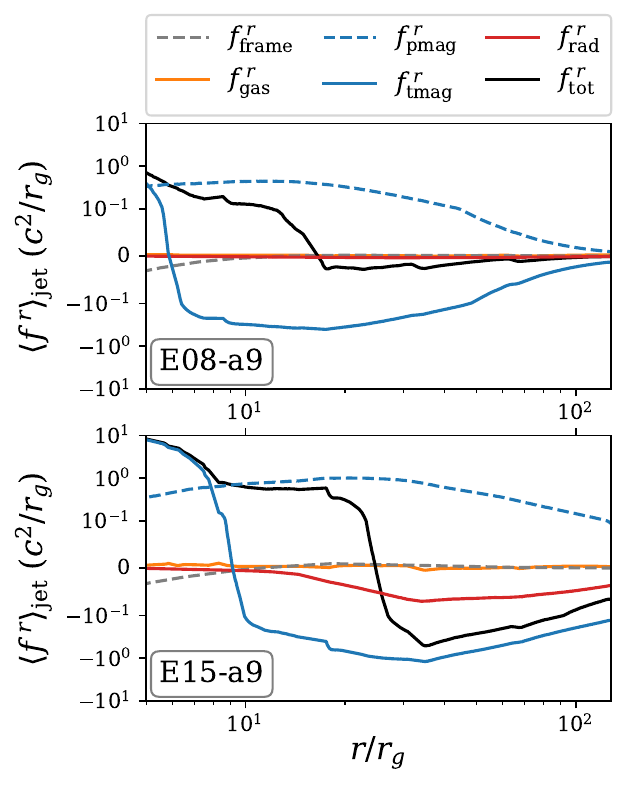}
    \caption{
    1D profiles of jet four-forces for the near-Eddington model E08-a9 (top) and the super-Eddington model E15-a9 (bottom), averaged over time and azimuth within the jet region.  The definitions of the four-force components are given in equation~(7) of \citetalias{PaperII}: frame force (gray dashed), gas pressure force (orange solid), magnetic pressure force (blue dashed), magnetic tension force (blue solid), and radiation force (red).  The total force is shown as a black solid line.  The derivation of these terms is provided in Appendix~D of \citetalias{PaperII}.  For reference, an additional super-Eddington model (E150-a9) is shown in Figure~12 of \citetalias{PaperII}.  
    }
    \label{fig:jet_force_compare}
\end{figure}

We analyze the four-force components to understand the physical origin of jet formation (see Section~3.6 in \citetalias{PaperII} for details).  The strong jet is initially launched by magnetic tension near the black hole.  This tension force is extremely powerful but acts only over a short range, accelerating the flow within a limited region ($\lesssim 10r_g$).  After this initial kick, the jet is further accelerated primarily by the magnetic pressure force until it reaches its peak velocity near $\sim20r_g$.  This acceleration depends critically on the effectiveness of inner disk collimation, which determines how strongly the magnetic pressure can be sustained over large distances.

\autoref{fig:jet_force_compare} compares the jet-averaged four-forces between the near-Eddington model E08-a9 and the super-Eddington model E15-a9.  Both models produce relativistic jets, but the jet in the near-Eddington case is less collimated due to its thinner inner disk.  The key difference arises from the ability to sustain magnetic pressure forces: stronger collimation confines the magnetic energy more effectively within a smaller solid angle, preventing its dilution and enhancing the energy gradient along the jet spine.  We also note that the radiation drag in model E15-a9 is much stronger than in model E08-a9, particularly at large radii, because the jet is accelerated to a much higher Lorentz factor. 

\begin{figure*}
    \centering
    \includegraphics[width=\textwidth]{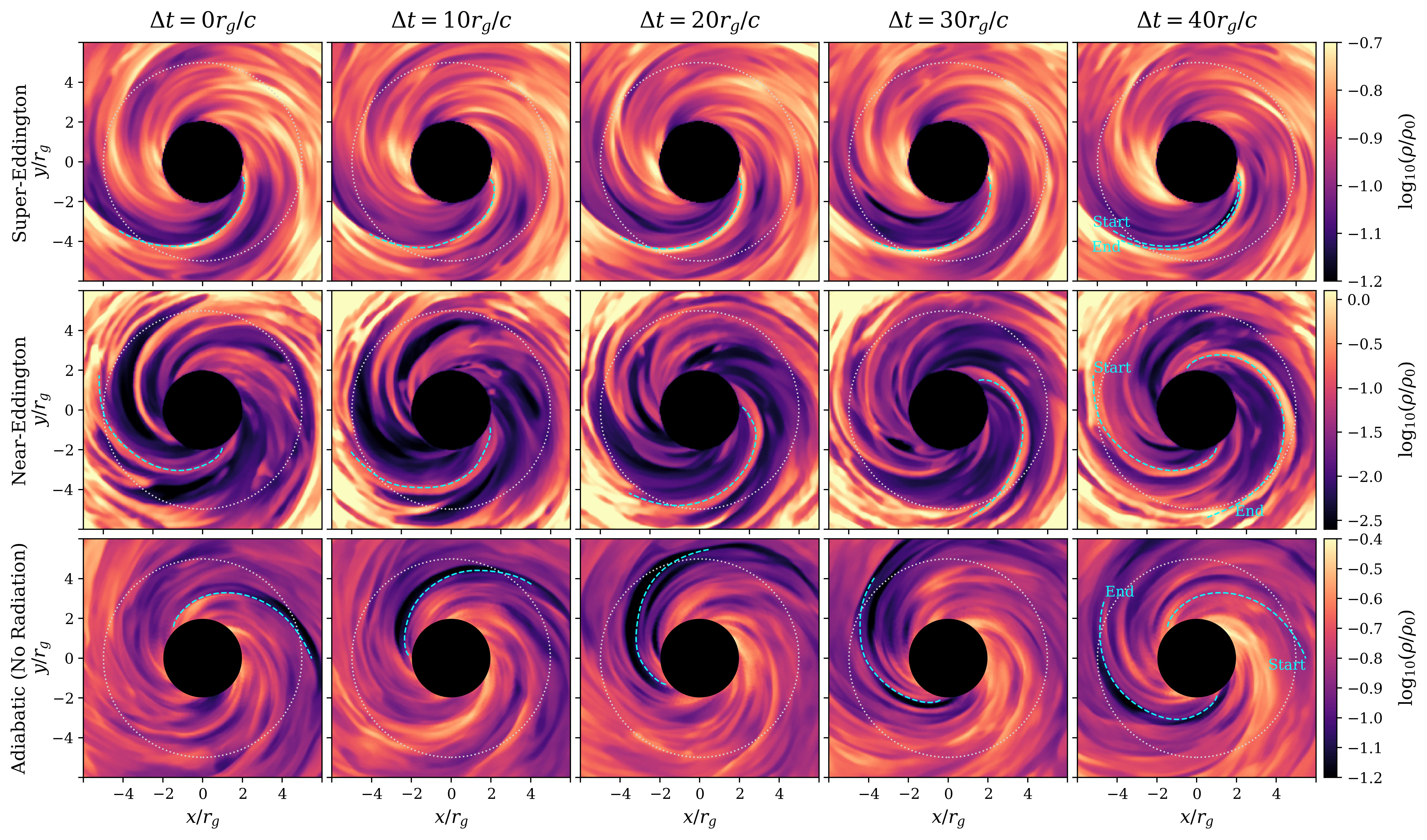}
    \caption{
    Spiral waves evolution near the black hole across different accretion regimes.  From top to bottom, we show representative super-Eddington (E88-a3), near-Eddington (E08-a3), and adiabatic (NoRad-a3) models.  From left to right, panels advance in time with a cadence of $\Delta t=10r_g/c$.  Each snapshot displays the midplane ($z=0$) density, with a tracked spiral arm (cyan dashed line) in either positive or negative perturbation.  The adiabatic model NoRad-a3 (bottom row) is non-radiative and thus corresponds to a very low accretion rate for which the flow becomes radiation-inefficient.  The starting snapshots are taken at $t=51800r_g/c$ for E88-a3, $t=57110r_g/c$ for E08-a3, and $t=32270r_g/c$ for NoRad-a3.  The grey dotted circle marks the ISCO.  In the final column, we also indicate the initial location of the tracked spiral arm for comparison. 
    }
    \label{fig:plunging}
\end{figure*}

\subsection{Spiral Waves in the Plunging Region}

Spiral density waves and shock-like spiral structures are commonly observed in non-axisymmetric accretion flows and have been studied both analytically and numerically in a variety of disk contexts \citep[e.g.,][]{Chakrabarti1990,Sawada1986,Savonije1994,Arzamasskiy2018,Aktar2021}.  In our simulations, the spiral structures are morphologically similar to those found in previous studies, while also being concentrated near the black hole and showing no clear transition at the innermost stable circular orbit (ISCO).  The phase relation and enhanced angular momentum extraction along the spiral arms are consistent with our earlier analysis of the super-Eddington models, as detailed in Section~3.8 of \citetalias{PaperII}. 

Building on this, we identify a positive correlation between the spiral wave phase velocity and the accretion regime.  Note that the non-radiative model corresponds physically to a very low accretion rate, placing the system in the radiation-inefficient (adiabatic) regime.  Tracking the spiral arms over time in \autoref{fig:plunging} reveals the following trends: (1) in the super-Eddington regime, the spiral wave behaves as a nearly standing, shock-like spiral density wave, with little noticeable motion; (2) in the near-Eddington regime, the spiral arm rotates around the black hole with a phase velocity substantially higher than in the super-Eddington case; and (3) in the adiabatic regime, the arm rotates at a similar, though slightly higher phase velocity compared with the near-Eddington model.

This systematic variation indicates that the plunging region dynamics cannot be determined solely by the spacetime metric; otherwise, the spiral wave behavior would be identical across regimes.  Although the level of radiation intensity differs significantly among these regimes, the radiation stress in the plunging region is always negligible, making radiation an unlikely driver of the observed trend.  

Maxwell stress, however, dominates the accretion process near the black hole, and the radial Maxwell stress (in simulation units) decreases with increasing accretion rate (e.g., NoRad-a3: $\sim 10^{-3}$, E08-a3: $\sim 6\times10^{-4}$, and E88-a3: $\sim 2\times10^{-4}$).  

Despite this correlation, the underlying physical cause remains uncertain.  Several factors complicate interpretation: (1) the disk structures differ substantially across regimes (thick vs. thin), as do their magnetic properties (turbulent vs. mean-field); and (2) the spiral wave properties may be influenced by fluid compressibility, which varies with the degree of radiation dominance.  Given these complexities, the origin of the phase velocity trend remains an open question that requires further investigation. 

\subsection{Comparison with Other Models}
\label{sec:compare_other_models}

We present a brief comparison with previous models in Section~3.6 of \citetalias{PaperI}.  Here, we focus on the near-Eddington regime and provide a more detailed discussion. 

\subsubsection{Newtonian MHD Models with Effective Cooling}

The envelope accretion in the thin thermal disk solution is similar to that found in earlier MHD models by \citet{Zhu2018} and \citet{Mishra2020}, respectively.  In both studies, accretion occurs mostly in the magnetic envelope and is driven by mean magnetic field torques, while a backward flow develops near the midplane.  This accretion pattern is highly consistent with all of our near- and sub-Eddington models in the presence of sufficient vertical magnetic flux.  

To achieve the thin disk regime, both \citet{Zhu2018} and \citet{Mishra2020} adopt effective cooling prescriptions by assuming an isothermal conditions.  In the absence of radiation pressure feedback, the reproduction of envelope accretion implies that this solution is not driven by radiation forces. 

Despite these similarities, the physical regime of the thermal disk is different.  In their simulations, the midplane remains gas pressure dominated and the accretion system corresponds to a low sub-Eddington regime, with accretion rates $\lesssim0.01\dot{M}_{\mathrm{Edd}}$.  The isothermal condition allows to produce a geometrically thin disk while maintaining sufficient thickness to resolve the MRI.  This regime is distinct from, yet complementary to, our radiation-dominated systems, which typically have accretion rates $\gtrsim0.01\dot{M}_{\mathrm{Edd}}$. 

\subsubsection{Newtonian MHD Models with Full Radiation Transport}

\citet{Huang2023} present three accretion models in a comparable accretion regime, solving full radiation transport within a Newtonian MHD framework.  Two of their models (XRB0.9 and XRB0.8) lie in the near-Eddington regime but differ in their initial magnetic topology.  The multiple-loop model XRB0.8 minimizes the net vertical magnetic flux and develops a fully magnetically elevated disk, closely aligned with our double-loop model E09-a3-DL.  

However, the single-loop model XRB0.9 evolves toward a steady-state solution that is qualitatively different from our single-loop models.  Model XRB0.9 exhibits a single-zone vertical structure in which magnetic and radiation pressures are comparable throughout the disk body and peak near the midplane.  This structure closely resembles the magnetically elevated disk solution (e.g., thier model XRB0.8 and our model E09-a3-DL), and is remarkably distinct from the thin disk solution with envelope accretion found in our models, as well as in \citet{Zhu2018} and \citet{Mishra2020}. 

Since \citet{Zhu2018} and \citet{Mishra2020} reproduce an envelope accretion pattern similar to ours using Newtonian MHD, this discrepancy is unlikely to originate from relativistic effects.  Instead, it likely reflects differences in the efficiency of vertical magnetic flux feeding onto the black hole.  In \citet{Huang2023}, the initial torus is located far from the accretor; during the relaxation phase, MRI-driven accretion transports material inward while gradually stretching the global magnetic field in the horizontal direction, leading to a slow accumulation of vertical magnetic flux.  In contrast, in \citet{Zhu2018} and \citet{Mishra2020}, the initial torus is placed closer to the black hole, allowing more rapid accumulation of vertical magnetic flux and facilitating the formation of a thin disk solution.  

These comparisons indicate that a single-loop magnetic configuration alone does not guarantee efficient feeding of vertical magnetic flux.  Instead, the effective magnetic topology that develops once the inner torus begins to interact with the accretor ultimately determines the resulting accretion flow structure.  Supporting this interpretation, a sub-Eddington model in \citet{Jiang2019}, which uses the same code and adopts a similar numerical setup, exhibits a solution comparable to model XRB0.9 at early times but later collapses into a thinner disk (see their Figure~2), likely as a consequence of continued vertical magnetic flux accumulation.  

These results suggest that magnetic envelope accretion requires the joint action of three key ingredients: (1) a geometrically thin disk enabled by efficient radiative cooling, (2) a sustained supply of vertical magnetic flux to the accretor (the source), and (3) efficient retention of magnetic flux at the accretor (the sink).  Beyond these factors, the intrinsic behavior of the magnetic field (dominated by turbulence or mean field) across different accretion regimes likely plays a crucial role in shaping the accretion dynamics, a topic we will further discuss in a separate paper. 

\subsubsection{GRMHD Models with M1 Radiation Transport}

In this section, we present a detailed comparison with previous radiation GRMHD models that employ the M1 approximation for radiation transport. 

\citet{Fragile2025} recently reported a series of radiation GRMHD simulations designed to explore the super-Eddington accretion regime.  Their models are initialized from an analytical solution with an imposed double-loop magnetic topology; however, all simulations ultimately evolve into the near-Eddington regime.  Since the initial conditions are constructed assuming a prescribed effective viscosity, these near-Eddington outcomes likely reflect the subsequent readjustment of the Maxwell stress by the self-consistently generated MRI during the system's evolution.

This readjustment naturally leads to an effective viscosity that differs from the initial assumption, suggesting that the final accretion rates are not tightly constrained by the disk profiles initialized under the assumed viscosity.  As a result, the near-Eddington accretion rates obtained in these models may mostly reflect the choice of initial conditions. 

Earlier pioneering work was carried out by S{\k{a}}dowski et al. \citep{Sadowski2016,Sadowski2016b,Lancova2019}.  Because these studies adopt a different radiative efficiency ($\eta=0.057$) from ours ($\eta=0.1$) in defining the Eddington accretion rate, their sub-Eddington regime corresponds closely to our near-Eddington regime.  As a result, a direct comparison between their simulations and ours is well motivated. 

In \citet{Sadowski2016}, two magnetic topologies are considered for comparison: a dipolar configuration (model D) and a quadrupolar configuration (model Q), analogous to our single-loop and double-loop magnetic setups, respectively.  The results are qualitatively consistent with our findings.  

Model Q develops a magnetically elevated disk, with magnetic pressure providing up to 80\% of the vertical support near the disk scale height.  This magnetic support helps stabilize the disk and suppress thermal instability (see the discussion in Section~2 of \citealt{Sadowski2016}).  Radiative cooling proceeds primarily through diffusion, and the effective viscosity is measured to be $\alpha\simeq0.09$ at $15r_g$, closely matching the values found in our double-loop models E09-a3-DL ($\alpha\simeq0.12$) and E07-a3-DL ($\alpha\simeq0.09$). 

Model D, by contrast, undergoes disk collapse, with the density becoming highly concentrated toward the midplane.  However, this solution is not well resolved due to computational limitations.  Based on these results, \citet{Sadowski2016} suggested that only model Q can sustain this accretion regime. 

The analysis in \citet{Sadowski2016} primarily focuses on model Q.  We note that model Q appears to exhibit magnetic polarity breaking in the polar regions, where a net vertical magnetic flux threads the horizon (see the second-to-last panel of Figure~8 in \citealt{Sadowski2016}).  Within the disk body, although the magnetic field topology is largely preserved and shows no net vertical flux, a small vortex-like magnetic structure develops near the midplane at $x\simeq-17r_g$.  This feature qualitatively resembles the magnetic structures observed during the transitional stages of our model E07-a3-DL (see the early and intermediate epochs in \autoref{fig:trans}).  

This raises the possibility that, if model Q were evolved for a longer duration, continued accumulation of vertical magnetic flux could eventually drive the system toward disk collapse.  Moreover, model~Q shows strong thermal support near the midplane (see Figure~9 of \citealt{Sadowski2016}), which is consistent with the intermediate epoch of our model E07-a3-DL.  These similarities suggest that model~Q may be in a transitional phase toward the thin thermal disk configuration, analogous to that observed in our model E07-a3-DL.   

In \citet{Lancova2019}, a follow-up study of model Q from \citet{Sadowski2016} shows that a steady ``puffy disk'' solution is established at an accretion rate of $1.05\dot{M}_{\mathrm{Edd}}$, corresponding to the near-Eddington regime under our definition.  Interestingly, despite being initialized without net vertical magnetic flux, this model evolves into a solution exhibiting magnetic envelope accretion with an embedded thin thermal disk.  This behavior closely resembles that of our transitional near-Eddington model E07-a3-DL, in which net vertical magnetic flux is gradually accumulated through radiation feedback (see \autoref{sec:magnetic_polarity_breaking} for details).

\subsubsection{Non-Radiative and Sub-Eddington Models}

The non-radiative models are effectively adiabatic and produce geometrically thick disks, in which MRI-driven turbulence locally transports angular momentum outward and dissipates energy.  This disk solution closely resembles that found in super-Eddington models, where photon trapping and low radiative efficiency render the accretion flow approximately adiabatic (see \citetalias{PaperII} for details).  These turbulence-dominated thick disks are fundamentally distinct from the near-Eddington solutions, in which mean magnetic fields significantly change the dynamics. 

In our sub-Eddington models, we likewise observe disk solutions that depend on the magnetic field topology; however, stronger radiative cooling produces geometrically thinner disks, and the simulations are therefore performed at higher grid resolution.  In follow-up papers, we will present a detailed analysis of these models, including the structure of midplane current sheets, and provide a comparison with the classical $\alpha$-disk model. 

\subsection{Observational Implications}
\label{sec:observational_implications}

The observational applications of our accretion models across different regimes are discussed in detail in \citetalias{PaperI}.  Here, we provide additional insight into how viewing-angle effects complicate the interpretation of observations of near-Eddington accretion systems, and we examine possible connections to ultrafast outflows as well as constraints from recent X-ray polarization measurements. 

\subsubsection{Apparent Luminosity and Viewing-Angle Effects}
As discussed in Section~4.2 of \citetalias{PaperI}, the outgoing radiation in the super- and near-Eddington models is strongly beamed.  We define the observational apparent luminosity as 
\begin{equation}
    L_{\mathrm{app}} =
    \frac
    {4\pi r^2\displaystyle \int_0^{2\pi}\int_{\theta_l}^{\theta_{l+1}}\langle-R^r_{\ t}\rangle_{\phi,t}\sqrt{-g} d\theta d\phi}
    {\displaystyle\int_0^{2\pi}\int_{\theta_l}^{\theta_{l+1}}\sqrt{-g} d\theta d\phi}
    \ .
    \label{eq:lum_app}
\end{equation}
This definition assumes an isotropically radiating source inferred from the observed flux at a given viewing angle. 

\begin{figure}
    \centering
    \includegraphics[width=\columnwidth]{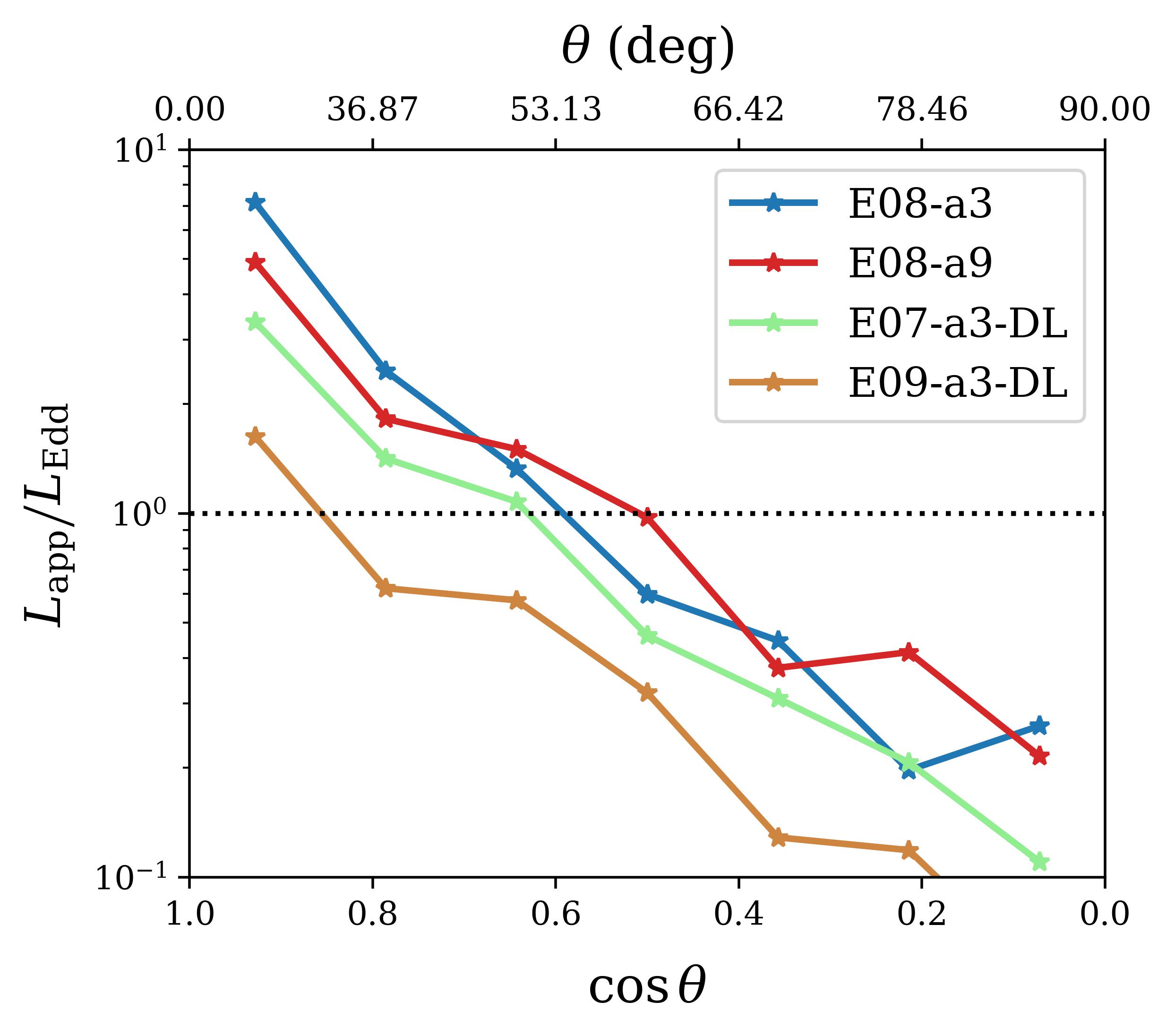}
    \caption{
    Apparent luminosity as a function of viewing angle for all near-Eddington models.  The measurement is taken at $r=800r_g$, following the definition of apparent luminosity in equation~\autoref{eq:lum_app}.  The radiation field is strongly beamed toward small inclination angles, causing the inferred luminosity to span nearly two orders of magnitude, from $\sim 0.1$ to $10 L_{\mathrm{Edd}}$, despite the underlying mass accretion rate being near Eddington. 
    }
    \label{fig:lum_app}
\end{figure}

\autoref{fig:lum_app} present the apparent luminosity at different viewing angles.  We use seven angular bins that evenly divide the solid angle to average the outgoing radial radiation flux and reduce ray effects, following Section~4.2 of \citetalias{PaperI}.  For the same near-Eddington accretion system, the inferred luminosity spans nearly two orders of magnitude, from $0.1$ to nearly $10L_{\mathrm{Edd}}$.  This range overlaps the low-luminosity end of ultraluminos X-ray sources (e.g., NGC~55 ULX-1) and the high-luminosity end of black hole X-ray binaries (e.g., GX~339-4).  Indeed, despite their different source classifications, their spectra (e.g., \citealt{Barra2022,Tamura2012}) exhibit similar shapes, characterized by a soft thermal component peaking at $1-3$~keV followed by a hard power-law tail. 

Furthermore, as the accretion rate increases from the near- to the super-Eddington regime, the radiative efficiency decreases significantly (e.g., Figure~9 of \citetalias{PaperI}).  This reduction can slow the growth of the observed luminosity and thereby complicate efforts to distinguish near- and super-Eddington accretion based solely on luminosity.  Since most black hole ULXs are believed to accrete in the mildly to moderately super-Eddington regime, spectral properties therefore provide a more reliable diagnostic of the accretion state.  

For example, Figure~12 of \citetalias{PaperI} shows that the emergent radiation in a super-Eddington model (E9-a3) is softer than that in a near-Eddington model (E08-a3).  This spectral softening primarily reflects a transition in disk type: near-Eddington systems develop a magnetic envelope that contributes a harder spectral component, whereas super-Eddington flows become fully radiation-supported and more strongly thermalized, resulting in a softer emergent spectrum.  A more quantitative diagnosis will require detailed spectral post-processing that incorporates viewing-angle effects (e.g., the increasing anisotropic beaming from the near- to super-Eddington regimes) in future studies. 

\subsubsection{Ultrafast Outflows, Disk Winds and Weak Jets}

The hot, mildly relativistic winds and weak jets present in our near-Eddington models may be relevant to the ultrafast outflows (UFOs) observed in luminous accreting black holes.  UFOs are commonly detected through highly blueshifted Fe-K absorption lines with velocities $\sim 0.1c$ in AGN \citep[e.g.,][]{Tombesi2010}, and have also been reported in ultraluminous X-ray sources \citep[e.g.,][]{Walton2016}.  In our simulations, the velocities of the winds and weak jets near the black hole reach comparable values (see \autoref{fig:outflow1d}, decreasing from $\sim 0.6c$ at $r=10r_g$ to $\sim0.01c$ at $r=300r_g$) and are driven by a combination of magnetic and radiative forces. 

Although the dominant driving mechanism of UFOs remains debated, particularly in sub-Eddington AGN where radiation pressure alone may be insufficient, our results demonstrate that hybrid MHD-radiative driving can naturally produce mildly relativistic outflows in near-Eddington systems.  Moreover, because near-Eddington models exhibit strong viewing-angle dependence, such systems may appear observationally sub-Eddington or even super-Eddington despite having intrinsically near-Eddington accretion rates (see \autoref{fig:lum_app}). 

In future work, we will extend our investigation by incorporating our sub-Eddington models to (1) measure wind column densities for direct observational comparison and (2) distinguish the respective roles of magnetic fields and radiation across accretion regimes.

\subsubsection{X-ray Polarization and Flow Geometry}

The black hole X-ray binary 4U~1630-47 was recently observed by IXPE during its high-soft state.  The measured polarization angle is aligned with the disk plane, indicating a scattering geometry extended parallel to the disk surface (e.g., a corona, atmosphere, or wind).  The polarization degree exceeds the prediction of the standard thin disk model, implying the presence of a geometrically thicker or vertically extended scattering layer.  These properties have been interpreted as consistent with a thin disk covered by a partially ionized atmosphere flowing outward at mildly relativistic velocities \citep{Ratheesh2024}.  This geometric configuration is broadly consistent with both classes of solutions in our near- and sub-Eddington models, which naturally produce vertically extended, moderately optically thin scattering layers above the disk surface. 

Moreover, X-ray polarization measurements can also constrain both the strength and geometry of global magnetic fields.  If magnetic fields near or above the scattering photosphere are sufficiently strong and ordered, Faraday rotation can significantly depolarize the emergent radiation \citep{Barnier2024}.  The observed high polarization degree may therefore place upper limits on the field strength and coherence scale in the emitting region.  However, alternative analyses suggest that Faraday effects may be modest within the IXPE band \citep{Krawczynski2026}. 

In future work, we will quantify Faraday rotation in both our near- and sub-Eddington models to further assess this issue.  We will also examine whether the magnetically elevated and thin thermal disk configurations produce distinguishable polarization signatures across accretion regimes.  If so, such differences could provide a direct observational diagnostic to discriminate between these two configurations.

\section{Conclusions} 
\label{sec:conclusions}

We carry out a parameter survey of black hole accretion in the radiation-dominated regime.  \citetalias{PaperI} provides an overview of this survey, while \citetalias{PaperII} establishes the analysis framework and applies it to the super-Eddington models.  This third paper focuses on the near-Eddington regime, yielding solutions that are significantly different from the super-Eddington cases. 

Our main results for the near-Eddington models are as follows: 
\begin{itemize}
    \item \textbf{Two disk solutions}: 
    We identify two stable accretion solutions in the near-Eddington regime: (1) a thin thermal disk embedded within a magnetic envelope, and (2) a magnetically elevated disk.  Which solution is realized depends sensitively on the availability of vertical magnetic flux; when sufficient vertical flux is present, the thin thermal disk favored. 
    
    \item \textbf{Vertical support}: 
    These two solutions exhibit distinct modes of vertical support.  (1) In thin thermal disks, the midplane is supported by gas and radiation pressure, while magnetic pressure compresses the structure; the overlying magnetic envelope is supported primarily by magnetic forces.  (2) In the magnetically elevated disk, vertical support is magnetic-dominated throughout the disk body, with radiation playing a subdominant role. 

    \item \textbf{Angular momentum transport}: 
    Near-Eddington accretion is strongly facilitated by mean-field Maxwell stresses. (1) In thin thermal disks, accretion proceeds primarily through the magnetic envelope.  (2) In magnetically elevated disks, accretion occurs throughout the disk body, with turbulent stresses contributing at a comparable level. 

    \item \textbf{Radiative cooling}: 
    Near-Eddington disks are moderately optically thick and can be efficiently cooled by radiative diffusion.  Radiative cooling is regulated by accretion heating (source) and radiative diffusion (sink).  In thin thermal disks, photons are produced near the midplane current sheet and energized via Comptonization in the overlying magnetic envelope, whereas in the magnetically elevated disk these processes occur more uniformly throughout the disk body.  
    
    \item \textbf{Outflows}: 
    Near-Eddington disks launch optically thin winds from the disk surface, driven by a combination of radiation and magnetic forces, and can also produce jets.  Both winds and jets become stronger at higher vertical magnetic flux; however, their efficiencies remain well below the radiative efficiency, in contrast to the super-Eddington models (see \autoref{sec:outflows_and_jet} for details). 
    

    
    \item \textbf{Magnetic flux evolution}: 
    At sufficiently high accretion rates, a weak or zero net vertical magnetic flux is difficult to sustain, as radiation-driven outflows can stochastically carry away unequal amounts of magnetic flux from the northern and southern disk surfaces.  This polarity breaking leads to a net accumulation of vertical magnetic flux in the inner disk, ultimately driving the system toward a thin thermal disk. 

    \item \textbf{Spin trends}: 
    In the models considered here, black hole spin influences both outflow efficiency and angular momentum transport.  
    The higher-spin models produce stronger outflows and less efficient angular momentum transport near the black hole.  In near-Eddington systems, spin effects extend to larger radii than in super-Eddington flows, likely reflecting the dominance of mean magnetic fields rather than turbulent stresses.  
    These results suggest that geometrically thin, radiation-efficient disks in the near- and sub-Eddington regimes are more likely to imprint spin-dependent features on the emerging radiation. 

    
\end{itemize}

\newpage
\begin{acknowledgments}
We thank Omer Blaes, Eliot Quataert, Sihao Cheng, Muni Zhou, Zhaohuan Zhu, Andrew Mummury, Chris Fragile, and Debora Lan\v{c}ov\'a for useful discussions.  This work was supported by the Schmidt Futures Fund, NASA TCAN grant 80NSSC21K0496, and the Simons Foundation.  The analysis made significant use of the following packages: NumPy \citep{Harris2020}, SciPy \citep{Virtanen2020}, and Matplotlib \citep{Hunter2007}.

An award for computer time was provided by the U.S. Department of Energy's (DOE) Innovative and Novel Computational Impact on Theory and Experiment (INCITE) Program. This research used supporting resources at the Argonne and the Oak Ridge Leadership Computing Facilities. The Argonne Leadership Computing Facility at Argonne National Laboratory is supported by the Office of Science of the U.S. DOE under Contract No. DE-AC02-06CH11357. The Oak Ridge Leadership Computing Facility at the Oak Ridge National Laboratory is supported by the Office of Science of the U.S. DOE under Contract No. DE-AC05-00OR22725. We thank Vassilios Mewes and Kyle Felker for support on these facilities.

Research presented in this article was supported by the Laboratory Directed Research and Development program of Los Alamos National Laboratory under project number 20220087DR.

This work has been assigned a document release number LA-UR-26-21541. 
\end{acknowledgments}

\appendix

\section{Simulation Quality Check} 

\begin{figure*}
    \centering
    \includegraphics[width=\textwidth]{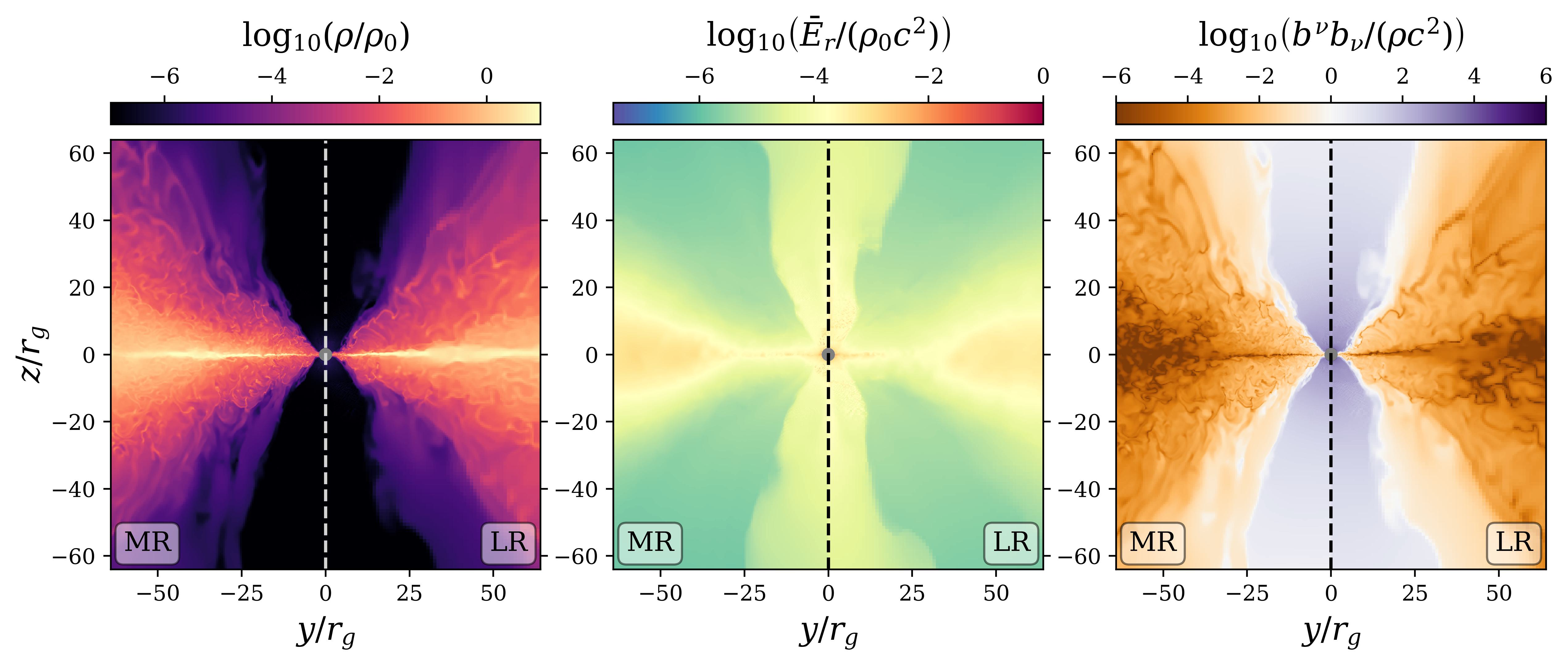}
    \caption{
    Comparison of model E08-a3 at different spatial resolutions. From left to right, the panels show density, fluid-frame radiation energy density, and magnetization at $t=60000r_g/c$, after the system has reached a steady state.  In each panel, the left half uses the intermediate-resolution (MR) grid and the right half uses the low-resolution (LR) grid.  The two resolutions produce qualitatively similar results, though the turbulent structure is more clearly resolved at higher resolution. 
    }
    \label{fig:res_compare}
\end{figure*}

\begin{figure*}
    \centering
    \includegraphics[width=\textwidth]{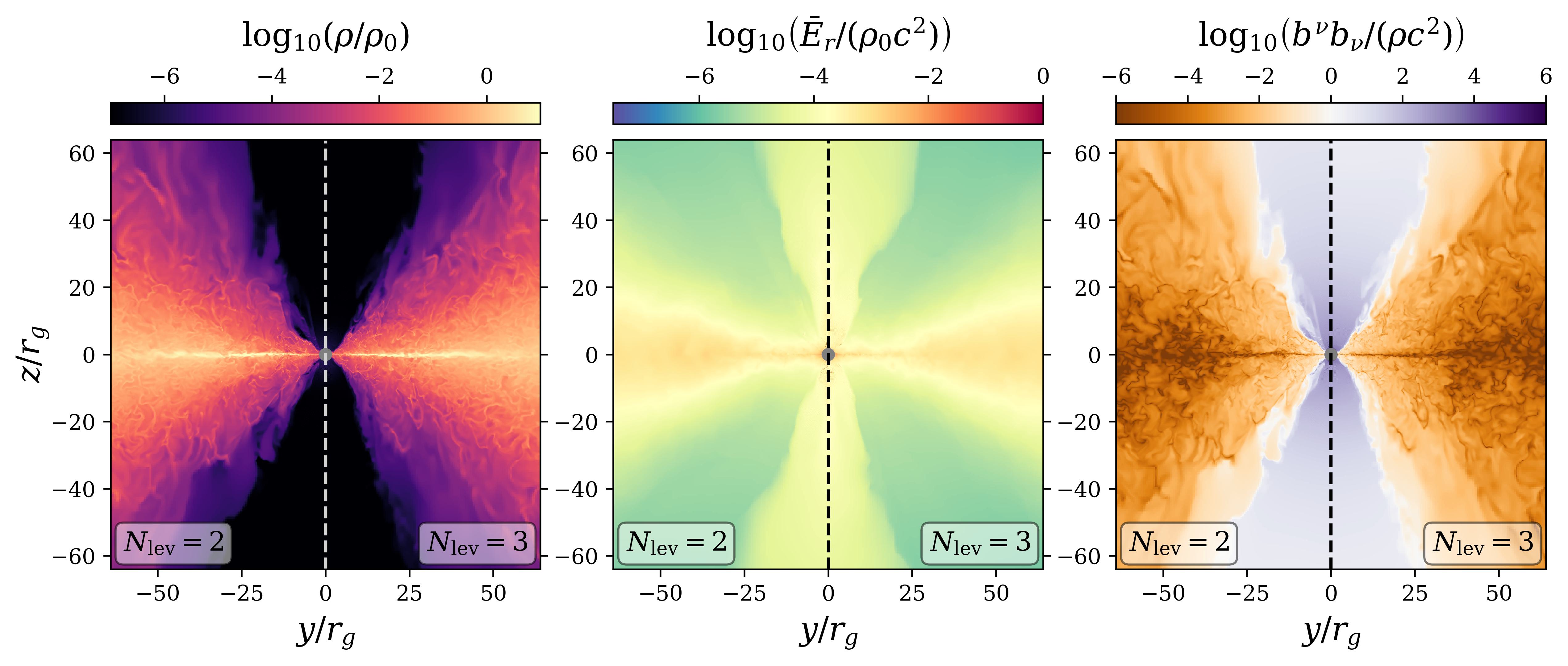}
    \caption{
    Comparison of model E08-a3 at different angular resolutions in radiation intensity. From left to right, the panels show the density, fluid-frame radiation energy density, and magnetization at $t=30000r_g/c$, after the system has reached a steady state.  In each panel, the left half shows the level-2 angular grid (42 angles) and the right half uses the level-3 angular grid (92 angles).  The two resolutions produce nearly identical structures, with the higher-resolution run exhibiting slightly sharper small-scale turbulent features near the jet boundary. 
    }
    \label{fig:res_compare_ang}
\end{figure*}

The magnetically elevated disk is generally easier to resolve with respect to MRI turbulence because its strong magnetic support yields relatively low gas density and high Alfv\'en speed.  A similar condition applies within the magnetic envelope of the thin thermal disk.  However, in the thermal midplane layer, the gas density becomes sharply concentrated, often reaching very high values.  The MRI quality factors then drop substantially, especially in the vertical direction, making thin thermal disk models computationally expensive.  Our sub-Eddington models (to be presented in Paper~IV) produce an even thinner disk and therefore employ higher resolution to adequately capture the dynamics near the midplane region. 

We adopt the same procedure for simulation quality assessment as in Section~4.2 of \citetalias{PaperII}.  \autoref{tab:quality} summarizes the quality factors and, where applicable, the resolution effects for all the near-Eddington models; all definitions follow \citetalias{PaperII}.  Measurements are taken at $10r_g$, representative of the inner disk and of the region where the intermediate- and low-resolution grids differ by a factor of two.  Resolution studies are presented only for the thin thermal disk models, as the magnetically elevated disk is already well resolved across the domain on the low-resolution grid. 

The disk-averaged quality factors exceed the criteria required to resolve both MRI turbulence \citep{Hawley2011, Hawley2013} and the thermal scale height for all near-Eddington models.  For the thin thermal disks, the factor-of-two change in resolution does not produce a correspondingly significant change in either the mass accretion rate or the outward angular momentum transport, indicating that these models are reasonably converged.  Interestingly, in the single-loop configurations, higher resolution in the low-spin case yields a slightly weaker accretion process, whereas the high-spin case shows the opposite trend.  

Note that for the thin thermal disk models, the disk-averaged MRI quality factors are dominated by contributions from the magnetic envelope.  We therefore also examine the quality factors within the thermal midplane layer.  In this region, the low-resolution models do not meet the standard MRI resolution criteria (e.g., $Q_z \gtrsim 6$, $Q_z \gtrsim 25$), and the intermediate-resolution models are only marginally resolved.  Specifically, the azimuthal quality factors remain sufficiently large, typically ranging from 30 to 200, whereas the vertical quality factors are marginal, with values of approximately 4-10. 

However, the outcomes from both resolution grids remain qualitatively (see \autoref{fig:res_compare}) and quantitatively (see \autoref{tab:quality}) consistent because accretion occurs primarily within the magnetic envelope, which is well resolved at both resolutions.  Consequently, the disk-averaged quality factors remain representative of the accretion process in the thin thermal disk.  We emphasize that commonly used MRI quality-factor criteria should be regarded as indicative rather than definitive, and their application requires careful judgment when assessing whether numerical models are adequately resolved.  Indeed, recent work by \citet{Jannaud2026} suggests that the MRI quality factor may not even be a reliable indicator of numerical convergence. 

We also assess the convergence of the radiation transport with respect to the angular discretization of the specific intensity. \autoref{fig:res_compare_ang} compares model E08-a3 evolved with the level-2 (42 angles) and level-3 (92 angles) angular grids.  The two runs produce nearly identical large-scale density, radiation, and magnetic structures, with the higher-angular-resolution model showing only slightly sharper small-scale turbulent features near the jet boundary, where radiation anisotropy is better resolved in the optically thin region.  This indicates that the level-2 angular grid already provides sufficient angular resolution for capturing the global radiation dynamics.

\begin{deluxetable*}{l c c c c c c c c c c c c}
\tablecaption{Simulation quality check \label{tab:quality}}
\tablehead{
    \colhead{Name} 
    & \colhead{$\dfrac{\big<\dot{M}_{10}\big>_{t}^{(\mathrm{MR})}}{\big<\dot{M}_{10}\big>_{t}^{(\mathrm{LR})}}$} 
    & \colhead{$\dfrac{\big<(T_{\mathrm{Rey}}^{\mathrm{turb}})^r_{\ \phi, 10}\big>_{\mathrm{disk}}^{(\mathrm{MR})}}{\big<(T_{\mathrm{Rey}}^{\mathrm{turb}})^r_{\ \phi, 10}\big>_{\mathrm{disk}}^{(\mathrm{LR})}}$}
    & \colhead{$\dfrac{\big<(T_{\mathrm{Max}})^r_{\ \phi, 10}\big>_{\mathrm{disk}}^{(\mathrm{MR})}}{\big<(T_{\mathrm{Max}})^r_{\ \phi, 10}\big>_{\mathrm{disk}}^{(\mathrm{LR})}}$}
    & \colhead{$\big<Q_{\mathrm{MRI,10}}^{\phi}\big>_{\mathrm{disk}}$}
    & \colhead{$\big<Q_{\mathrm{MRI,10}}^z\big>_{\mathrm{disk}}$}
    & \colhead{$\big<Q_{\mathrm{therm,10}}\big>_{\mathrm{disk}}$}
    \\
    \quad\;\;\;(1) & (2) & (3) & (4) & (5) & (6) & (7)
}
\startdata
    E09-a3-DL & -    & -    & -    & 201\qquad\:\: & 40\quad\:\: & 21\quad\:\:\:\: &  \\
    E08-a9    & 1.29 & 1.55 & 1.30 & 388 (221)     & 114 (65)    & 46 (25)         &  \\
    E08-a3    & 0.90 & 0.75 & 0.90 & 509 (216)     & 155 (71)    & 42 (18)         &  \\
    E07-a3-DL & 0.75 & 1.07 & 1.12 & 189 (123)     & \:\:37 (20) & 30 (21)         &  \\ 
    \hline
\enddata
\tablecomments{
    The subscript `10' indicates measurements at $10r_g$. Superscripts `(MR)' and `(LR)' refer to the intermediate- and low-resolution models, respectively.  Quality factors in parenthesis are measured from the low-resolution models.  All relevant definitions are provided in \citetalias{PaperII}. 
    {\bf Columns (from left to right):}
    (1)~Model name; 
    (2) Ratio of mass accretion rates (between intermediate- and low-resolution models); 
    (3) Ratio of turbulent Reynolds stress; 
    (4) Ratio of Maxwell stress; 
    (5) Azimuthal MRI quality factor; 
    (6) Vertical MRI quality factor; 
    (7) Thermal quality factor. 
    }
\end{deluxetable*}

\bibliography{edd_paper_iii}
\bibliographystyle{aasjournal}

\end{CJK*}
\end{document}